\documentclass[journal,twocolumn,final]{IEEEtran}

\usepackage{cite}
\usepackage{pbox}
\usepackage{psfrag}
\usepackage{color}
\usepackage{pstricks}
\usepackage{graphicx}
\usepackage{amsmath}
\usepackage{amssymb}
\usepackage{amsthm}
\usepackage{mathrsfs}
\usepackage{mathtools}
\usepackage{subcaption}
\usepackage{tikz,pgfplots}
\pgfplotsset{compat=newest}
\usetikzlibrary{graphs,matrix,positioning,calc,fit,backgrounds,chains}
\usepackage[breakable]{tcolorbox}
\tcbuselibrary{breakable}
\usepackage{tikzscale}

\let\oldbrace\{
\def\{{\oldbrace\kern0.5pt}

\def\tr{\mathop{\rm tr}\nolimits}%
\newcommand{\nn}{\nonumber}

\newcommand{\Ac}{\mathcal{A}}
\newcommand{\Bc}{\mathcal{B}}
\newcommand{\Cc}{\mathcal{C}}

\newcommand{\Ec}{\mathcal{E}}
\newcommand{\Fc}{\mathcal{F}}
\newcommand{\Gc}{\mathcal{G}}

\newcommand{\Nc}{\mathcal{N}}
\newcommand{\Oc}{\mathcal{O}}
\newcommand{\Pc}{\mathcal{P}}

\newcommand{\Sc}{\mathcal{S}}
\newcommand{\Tc}{\mathcal{T}}

\newcommand{\Xc}{\mathcal{X}}
\newcommand{\Yc}{\mathcal{Y}}

\newcommand{\Ds}{\mathsf{D}}

\newcommand{\av}{\boldsymbol{a}}

\newcommand{\Nv}{\boldsymbol{N}}

\newcommand{\Rv}{\boldsymbol{R}}
\newcommand{\Xv}{\boldsymbol{X}}
\newcommand{\Wv}{\boldsymbol{W}}
\newcommand{\Yv}{\boldsymbol{Y}}
\newcommand{\Zv}{\boldsymbol{Z}}

\newcommand{\xv}{\boldsymbol{x}}
\newcommand{\yv}{\boldsymbol{y}}

\newcommand{\vv}{\boldsymbol{v}}
\newcommand{\sv}{\boldsymbol{s}}

\newcommand{\pen}{{P_e^{(n)}}}

\newcommand{\aep}{{\mathcal{T}_{\epsilon}^{(n)}}}

\newcommand{\aepk}{{\mathcal{T}_{\epsilon}^{(k)}}}

\newcommand{\Sh}{{\hat{S}}}

\newcommand{\mh}{{\hat{m}}}
\newcommand{\sh}{{\hat{s}}}

\newcommand{\Yt}{{\tilde{Y}}}

\newcommand{\Ht}{{\tilde{H}}}

\newcommand{\yt}{{\tilde{y}}}

\def\d{\delta}
\def\e{\epsilon}

\DeclareMathOperator\E{\sf E}
\let\P\relax
\DeclareMathOperator\P{\sf P}

\DeclareMathOperator*{\argmax}{arg\,max}

\newtheorem{theorem}{Theorem}
\newtheorem{lemma}{Lemma}
\newtheorem{corollary}{Corollary}

\newtheorem{assumption}{Assumption}

\theoremstyle{definition}

\newtheorem{remark}{Remark}

\DeclareMathOperator*{\esssup}{\textrm{ess}\sup}
\DeclareMathOperator*{\esssupe}{\textrm{\em ess}\sup}

\newcommand{\FAR}{{\textsf{FAR}}}
\newcommand{\WADD}{\textsf{WADD}}
\newcommand{\SCS}{\textsf{SCS}}

\newcommand{\Ccn}{\Cc^{(n)}}
\newcommand{\SUM}{\textsf{SUM}}

\IEEEoverridecommandlockouts
\begin{document}
	\title{On the Fundamental Tradeoff of Joint Communication and QCD: The Monostatic Case}

	\author{Sung Hoon Lim and Daewon Seo
		\thanks{ Sung Hoon Lim is with the School of Information Sciences, Hallym University, Chuncheon 24252, South Korea (e-mail: shlim@hallym.ac.kr). D.~Seo is with the Department of Electrical Engineering and Computer Science, Daegu Gyeongbuk Institute of Science and Technology (DGIST), Daegu 42988, South Korea (e-mail: dwseo@dgist.ac.kr). }
		}

\maketitle
\allowdisplaybreaks

\begin{abstract}
This paper investigates the fundamental tradeoff between communication and quickest change detection (QCD) in integrated sensing and communication (ISAC) systems under a monostatic setup. We introduce a novel  Joint Communication and quickest Change subblock coding Strategy (JCCS) that leverages feedback to adapt coding dynamically based on real-time state estimation. The achievable rate-delay region is characterized using state-dependent mutual information and KL divergence, providing a comprehensive framework for analyzing the interplay between communication performance and detection delay. Moreover, we provide a partial converse demonstrating the asymptotic optimality of the proposed detection algorithm within the JCCS framework. To illustrate the practical implications, we analyze binary and MIMO Gaussian channels, revealing insights into achieving optimal tradeoffs in ISAC system design.
\end{abstract}

\begin{IEEEkeywords}
	Integrated sensing and communication, quickest change detection, CuSum test, constant subblock-composition codes
\end{IEEEkeywords}

\section{Introduction}

The ability to simultaneously perform communication and sensing tasks within shared resources is a defining feature of integrated sensing and communication (ISAC) systems, which are envisioned to play a central role in next-generation wireless networks~\cite{Bourdoux2020_2,Liu--Masouros--Petropulu--Griffiths--Hanzo2020}. Unlike traditional systems that dedicate separate resources and hardware to communication and sensing respectively, ISAC systems are designed to share spectrum, power, and hardware, thereby enhancing efficiency and reducing costs~\cite{Sturm--Wiesbeck2011,Liu--Cui--Masouros--Xu--Han--Eldar--Buzzi2022}. This unified approach enables ISAC to support a variety of emerging applications, such as device localization, autonomous navigation, surveillance, and healthcare~\cite{Akan--Arik2020,Liu--Huang--Li--wan--Li--Han--Liu--Du--Tan--Lu--Shen--Colone--Chetty2022}. 

In ISAC, a primary focus is the tradeoff between communication rate and sensing accuracy, which has been characterized in several scenarios. Early investigations on this tradeoff date back even before the formal emergence of ISAC~\cite{Sutivong--Chiang--Cover--Kim2005, Zhang--Vedantam--Mitra2011}. Since then, \cite{Joudeh--willems2022} initiated an information-theoretic study specifically for current form of ISAC. Also, \cite{Xiong--Liu--Cui--Yuan--Han--Caire2023} has examined the tradeoff between information rate and mean-squared error (MSE) in state estimation, while~\cite{Ahmadipour--Kobayashi--Wigger--Caire2024, Chang--Wang--Erdogan--Bloch--2023} have analyzed the relationship between communication rate and either detection error probability or distortion. Recent advancements include investigations into ISAC for multiple-input multiple-output (MIMO) systems~\cite{Xiong--Liu--Cui--Yuan--Han--Caire2023, Ouyang--Liu--Yang2023, Liu--Huang--Zheng2024, Smida--Alexandropoulos--Riihonen--Islam2024} and multiuser scenarios~\cite{Ahmadipour--Wigger2023, Liu--Wan--Yi--Qiu--Caire2024}. Moreover, machine learning methods have been integrated into ISAC design, as demonstrated in~\cite{Bian--Zhang--Gunduz2024, Nikbakht--Wigger--Shamai--Poor2024}, and studies for unequal error protection have been considered in~\cite{AhmadipourWS2024, Seo--Lim2024b, Wu--Joudeh2024}. Beyond these intrinsic ISAC metrics, secrecy constraints have also been incorporated, with studies such as~\cite{Gunlu--Bloch--Schaefer--Yener2023, Ren--Qiu--Xu--Ng2023} examining secure communication in the presence of simultaneous sensing tasks. Comprehensive surveys, including~\cite{Liu--Cui--Masouros--Xu--Han--Eldar--Buzzi2022, Liu--Huang--Li--wan--Li--Han--Liu--Du--Tan--Lu--Shen--Colone--Chetty2022, Cui--Liu--Jing--Mu2021, Zheng--Lops--Eldar--Wang2019}, offer broader perspectives on ISAC’s potential and challenges.

Building upon this foundation, we address the delay of sensing in a quickest change detection (QCD) framework. QCD problems aim to detect abrupt changes in statistical distributions as quickly as possible while minimizing false alarms~\cite{Lorden1971, Pollak1985, Lai1998, Veeravalli--Banerjee2014, Veeravalli--Fellouris--Moustakides2024}, which is a critical but less explored aspect of ISAC. As a representative example, in next-generation mobility systems such as autonomous vehicles and unmanned drones, minimizing sensing delay is essential for real-time decision making (e.g., obstacle detection) or reliable communication (e.g., channel tracking); incorporating the QCD perspective into ISAC enables faster sensing, thereby supporting low-latency decision making and robust communication in dynamic environments. Similar requirements arise in other latency-sensitive applications, including robotics, remote healthcare, surveillance, and defense~\cite{Xie--Zou--Xie--Veeravalli2021}.

On the sensing side, while traditional QCD studies optimize detection under sensing-only setups~\cite{Veeravalli--Fellouris--Moustakides2024}, our work incorporates a simultaneous communication requirement, significantly complicating the design of coding and detection strategies. On the communication side, investigations on state-dependent communication channels provide relevant theoretical insights. For instance,~\cite{Kim--Sutivong--Cover2008, Zhang--Vedantam--Mitra2011, Choudhuri--Kim--Mitra2020} explore state estimation and amplification in channels with state information. However, these studies usually focus on average detection performance, such as average detection accuracy, whereas QCD inherently demands a worst-case consideration of detection delay minimization. Moreover, the initial study of ISAC systems in the context of QCD was presented on a bistatic model which inherently utilizes open-loop coding strategies~\cite{Seo--Lim2024}. 
This work significantly extends our earlier results in~\cite{Seo--Lim2024} by addressing the monostatic setting. Whereas~\cite{Seo--Lim2024} focused on a bistatic architecture with non-adaptive (open-loop) coding, the present work considers a feedback-enabled scenario where the transmitter dynamically adapts its coding strategy based on feedback signals, i.e., delayed channel output observations. The resulting formulation poses a more challenging problem for characterizing the fundamental tradeoffs between communication rates and detection delays, particularly due to the presence of feedback-based adaptive coding.
The proposed strategy, termed the Joint Communication and quickest Change subblock coding Strategy (JCCS), is specifically designed to exploit the structure of the monostatic model. Technically, both works use constant subblock composition codes, but the present scheme incorporates dedicated pilot symbols to estimate the channel state, with the feedback signal induced by these pilots subsequently used to adaptively transmit state-dependent codewords. This estimation-and-adaptation mechanism introduces new statistical dependencies and is a key innovation absent from~\cite{Seo--Lim2024}.
On the detection side, we develop a Subblock CuSum (SCS) test that leverages the estimated state sequence and pilot-aided observations. While both papers rely on sequential probability ratio test principles, the current analysis generalizes to multi-state post-change models and provides cleaner, more refined detection guarantees under adaptive encoding.
Finally, although both works apply asymptotic techniques based on Lai’s quickest change detection framework, the converse analysis here differs from that of~\cite{Seo--Lim2024}, which closely follow and extent the proof of~\cite{Lai1998}. In particular, our converse explicitly incorporates the interaction between adaptive coding and state-aware detection. As a result, this paper offers a more comprehensive and technically challenging treatment that generalizes and strengthens our earlier results.

In the present work, we make the following key contributions:
\begin{enumerate}
    \item \textbf{Achievable tradeoff region:} We characterize an achievable rate-delay region for the joint communication and QCD setup for the monostatic model. The results are summarized in Thm.~\ref{thm:theorem1}, where achievable rates are expressed in terms of state-dependent mutual information, and delay constraints are tied to KL divergences under false alarm limitations.

    \item \textbf{Novel coding and detection strategies:} We develop feedback-based closed-loop coding strategies that exploit echo feedback for real-time state detection. Additionally, we propose a tailored detection mechanism based on a generalized cumulative sum (CuSum) test designed for the ISAC setting.

    \item \textbf{Converse result on detection optimality:} We establish a converse result in Thm.~\ref{thm:conv} proving the asymptotic optimality of our proposed detection algorithm within the framework of our coding strategy. This result ensures that the delay performance achieved by our method is fundamentally optimal for the given codebook design.
\end{enumerate}

The developed coding strategy overcomes several technical challenges. We build on the adaptive QCD strategy developed for the stand-alone QCD problem in~\cite{Veeravalli--Fellouris--Moustakides2024}. To integrate the strategy into a joint communication setting, we develop adaptive constant subblock composition codes. Additional pilot symbols are used to detect the state, and the codeword composition is chosen adaptively based on the state detection. Our strategy allows the codewords to adopt a symbol distribution that attains some tradeoff point for communication and QCD.

Also, we provide a partial converse showing that, under our proposed code construction, the Subblock CuSum (SCS) detection strategy is asymptotically optimal in the sense of minimizing the worst-case detection delay. While this result provides insight into the structure of good detection strategies under our framework, it is partial in the sense that it does not characterize the whole fundamental tradeoff between rate and delay. Rather, it establishes optimality of the QCD detection strategy under the specific codebook design adopted in this work.
By addressing the complexities of feedback-aided state-dependent channels, this work offers both theoretical foundations and practical guidelines for future ISAC systems. Notably, we demonstrate how adaptive strategies leveraging feedback expand the achievable rate-delay region, revealing new possibilities for system optimization. 

\textit{Notation:} We closely follow the notation in~\cite{El-Gamal--Kim2011}. Calligraphic letters denote alphabet spaces, and uppercase and lowercase letters respectively denote random variables and their realizations. 
For a sequence of random variables, we write $X_{i_1}^{i_2} = X_{i_1}, \ldots, X_{i_2}$, and in particular, $X^i = X_1,\ldots, X_i$. For a sequence $x^n$ on $\Xc^n$, the type or composition of $x^n$ is defined to be $\pi(x | x^n) := \big| \{ i \colon x_i = x \} \big| /n$ for $x \in \Xc$. Sequences are also written in vector notation, e.g., a length $L$ vector is denoted by $\xv$ where the length will be specified in the context. The notation for a sequence of vectors follow the same convention $\xv^k = \xv_1,\ldots, \xv_k$. 
The Kullback-Leibler (KL) divergence between two distributions $p_{Y|X}(y|x)$ and $q_{Y|X}(y|x)$ and its conditional version are denoted by
\begin{align*}
    D(p_{Y|X}\|q_{Y|X}|x) &= \sum_{y\in\Yc} p_{Y|X}(y|x)\log \frac{p_{Y|X}(y|x)}{q_{Y|X}(y|x)}, \\
    D(p_{Y|X}\|q_{Y|X}|p_X) &= \E_{p_X} [D(p_{Y|X}\|q_{Y|X}|X)].
\end{align*}

\section{Problem Statement} \label{sec:problem_statement}
\subsection{Channel model}

\begin{figure}[t]
	\centering
	\resizebox{!}{10em}{\begin{tikzpicture}[font=\large, node distance=.6cm and 1cm, start chain]
     \tikzstyle{rect}=[draw=black, 
                   rectangle, 
                   text opacity=1,
                   minimum width=50pt, 
                   minimum height = 25pt, 
                   align=center]
  \node[rect] (encoder) {Encoder};
  \node[rect, above=of encoder] (detector) {Detector};
  
  \node[rect, right=2.9cm of encoder] (chn) {$p_{Y,\tilde{Y}|X, S}$};
  \node[rect, right=of chn, label={below:Receiver}] (decoder) {$\hat m(\tilde y^n)$};

    \draw[-stealth] (encoder.east) node (line1)[right]{}-- +(0.2,0) |- (detector.east);
  \node[fit=(encoder)(detector)(line1), label={below:Transmitter}, draw, dashed] (transmitter) {};

  \node[rect, above right=2mm of transmitter] (delay) {delay};

\draw[-stealth] (encoder.east) -- node[above] {$~X_i(M, Y^{i-1})$} (chn.west);
\draw[-stealth] (detector.south) -- (encoder.north);
\draw[-stealth] (chn.north)  |- node[above left] {${Y}_{i}$}   (delay.east);
\draw[-stealth] (delay.west)  -| node[above ] {${Y}_{i-1}$}   (transmitter.north);
\draw[stealth-] (encoder.west) --+ (-5mm,0) node[left] {$M$};
\draw[-stealth] (detector.west) --+ (-5mm,0) node[left] {$N$};
\draw[-stealth] (chn.east) -- node[above] {$\Yt_i$}(decoder.west);
\draw[-stealth] (decoder.east) --+ (5mm,0) node[right] {$\hat{M}$};
\draw[stealth-] (chn.south) --+ (0, -5mm) node[below] {$S_i$};
\end{tikzpicture}}
	\caption{Monostatic model for ISAC. The encoder and QCD detector are the same entity which has access to the channel observation $Y$ via feedback.} 
	\label{fig:monostatic_model}
\end{figure}
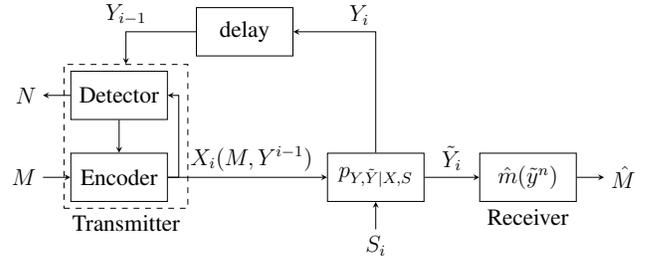

Consider a state-dependent memoryless broadcast channel $p_{Y,\Yt|X,S}(y,\yt|x, s)$ depicted in Fig.~\ref{fig:monostatic_model}. This channel can be expressed as the product of two conditionally independent components:
\begin{align}
    p(y^n, \yt^n | x^n, s^n) = \prod_{i=1}^n p(y_i | x_i, s_i) p(\yt_i | x_i, s_i), \label{eq:dmc}
\end{align}
where $s^n$ represents the state sequence, $x^n$ is the channel input, and $y^n, \yt^n$ are the channel output sequences. Specifically, $\yt^n$ is observed at the communication receiver to enable reliable decoding of the transmitted message, while $y^n$, representing the feedback (e.g., an echo signal), is observed at the transmitter and aids in state detection. 

The state sequence $s^n$ follows a deterministic and step-like constant pattern. Initially, the sequence remains in a base state, denoted by $s_i = 0$ for $i < \nu$, up to a specific (unknown) change point $\nu$. After $\nu$, the state transitions to a new unknown state $s \in \Sc$, where $\Sc = \{1, \ldots, |\Sc|\}$ is a finite set of possible post-change states. Formally, the state sequence is expressed as:
\begin{align*}
    s_i = \begin{cases}
        0 & \text{for } i < \nu, \\
        s & \text{for } i \ge \nu,
    \end{cases}
\end{align*}
where $s \neq 0$.

The feedback output $y_i$ serves a dual purpose. It supports state-aware message encoding, which enhances communication reliability over the state-dependent channel. Additionally, $y^n$ serves as the observation for quickest change detection (QCD) at the transmitter/detector. This dual 
objective is inherently coupled since accurate state detection enhances both QCD and communication performance.

The following assumptions are made in our channel model to ensure that the problem remains well-defined and analytically tractable. For simplicity, throughout the paper, we will often denote $p^{(s)} = p_{Y|X, s}$.

\begin{assumption}\label{assump: distinguishable_channels}
For every pair of states $s_1\neq s_2$, where 
$s_1\in\Sc$ and $s_2\in\Sc\cup\{0\}$, there exists an input $x\in\Xc$ such that
\begin{align}
    D(p^{(s_1)} \|p^{(s_2)} | x)>0. 
\end{align}
\end{assumption}
This assumption guarantees that for any pair of states $s_1$ and $s_2$, there exists at least one input $x$ that makes the output distributions distinguishable. Without this property, the problem would become degenerate, as state transitions would not be inferable through feedback. The next is a mild technical assumption additionally needed in our proof.
\begin{assumption}\label{assump: finite_distance}
    For every state $s\in\Sc$ and input $x\in\Xc$, the following holds:
    \begin{align}
        \sum_{y\in\Yc} p^{(s)}(y|x) \left(\log\frac{p^{(s)}(y|x)}{p^{(0)}(y|x)}\right)^2  < V <\infty, \label{eq:finite_V}
    \end{align}
where $V$ is a finite constant.
\end{assumption}

\subsection{Code definition}
A code in this work simultaneously serves two purposes: 1) it functions as a conventional channel code to enable reliable transmission of messages over the state-dependent noisy channel $p_{\Yt|X,S}$, and 2) it acts as an active sensing signal to identify the unknown change point $\nu$ as quick as possible from the feedback observations through the channel $p_{Y|X,S}$. 

Formally, a code for joint communication and QCD has the following components. For a given block length $n$, an $(2^{nR}, n)$ code consists of
\begin{itemize}
	\item a message set $m \in [1:2^{nR}]$,
	\item an encoder that assigns a symbol $x_i(m, y^{i-1})$, $i\in[1:n]$ to each message $m\in[1:2^{nR}]$ and past observation $y^{i-1}\in\Yc^{i-1}$,
	\item a decoder that assigns a message estimate $\hat{m}(\yt^n)$ to each observation sequence $\yt^n$, and
	\item a stopping rule $N(m)$ such that the event $\{N=i\}$ is measurable with respect to $Y^i$ and $x_1(m), x_2(m, Y_2),\ldots, x_i(m,Y^{i-1})$. If the stopping rule is not activated until the end of the codeword, we define $N = n+1$. 
\end{itemize}

\subsection{Performance metrics}
Next, we define the performance metrics associated with our problem. For communication, the performance metric is the average probability of error, denoted by
$\pen = \P(\hat{M} \neq M)$,
where $\hat{M}$ is the decoder's estimate. A rate $R \in \mathbb{R}_+$ is said to be \textit{achievable} if there exists a sequence of codes such that the probability of error $\pen \to 0$ as the block length $n \to \infty$.

For the QCD metrics, we extend Lorden's formulation~\cite{Lorden1971} to the ISAC setting.
The false alarm rate (\FAR) for a given codebook is defined as
\begin{align}
	\FAR(N) &=  \limsup_{n\to\infty}\max_{m\in[1:2^{nR}]} \frac{1}{\E_{\infty}[ N(m) ]}, \label{eq:cond_FAR_max}
\end{align}
where $\E_{\infty}$ denotes the expectation under the assumption $\nu=\infty$, i.e., the transmission is entirely over the base-state channel $p^{(0)}$.
This definition resembles the bistatic formulation in~\cite{Seo--Lim2024} but incorporates the randomness of $Y^{i-1}$ through feedback adaptive coding. 

Next, the worst average detection delay (\textsf{WADD}) objective is defined for the worst possible realizations $\nu$ under post-change state $s \in \Sc$ (recall that $\Sc$ excludes the base state), as

\begin{align}
\WADD(N,s) &= \bar\E^{(s)}_{\nu}[(N(m)-\nu+1)^+ | \Fc_{\nu-1}(m)] \label{eq:ps_wadd_max}
\end{align}
where 
\begin{align*}
\bar \E^{(s)}_{\nu}[\cdot] = \sup_{\nu \ge 1 } \limsup_{n\to\infty} \max_{m\in[1:2^{nR}]} \esssup \E^{(s)}_{\nu}[\cdot],    
\end{align*}
$\Fc_i(m) = \sigma(Y^i, X_1(m), X_2(m,Y_1),\ldots, X_i(m, Y^{i-1}))$
is the filtration generated by the observation and codeword sequence up to time $i$, and the essential supremum is taken with respect to the filtration $\Fc_{\nu-1}(m)$. Here, $\E_\nu^{(s)}$ denotes the expectation under the assumption that the state transition occurs at time $\nu$ to state $s$.

To minimize the delay under an unknown change point, we define the set of delay constraints $\Ds = \{\Ds_1,\ldots, \Ds_{|\Sc|}\}$ and say that a {\WADD} `$\Ds$' under a {\FAR} constraint `$\alpha$' is achievable if there exists a sequence of codebooks $\mathcal{C}^{(n)}$ and stopping rules $N(m)$ such that 
\begin{align*}
	\inf_{N(m), \{\Ccn\}: \FAR(N) \le \alpha} \WADD(N,s) \le \Ds_s, 
\end{align*}
for all $s\in\Sc$.
To succinctly present our results, for each $s\in\Sc$, we define the ``speed'' of detection
\begin{align}
	\Delta_s := \lim_{\alpha \to 0} \frac{|\log \alpha|}{\Ds_s}\label{eq:Delta}
\end{align}
and say that $\Delta=(\Delta_1,\ldots, \Delta_{|\Sc|})$ is achievable if there exists a pair $(\alpha, \Ds)$ that asymptotically attains $\Delta$ as $\alpha \to 0$. The goal is to characterize the set of achievable $(R, \Delta)$ pairs, namely, the $R$-$\Delta$ region:
\begin{align}
	\mathscr{R} &= \text{closure} \{ (R, \Delta ): \text{$R$ and $\Delta$ are achievable} \}. \label{eq:region}
\end{align}

\subsection{Supplementary definitions}
We introduce several technical definitions that are used throughout this work. For any two states $s_1, s_2 \in \Sc$, the Bhattacharya coefficient for input $x$ is defined as
\begin{align*}
    \rho(s_1, s_2|x) := \sum_y \sqrt{ p^{(s_1)}(y|x) p^{(s_2)}(y|x)},
\end{align*}
where $p^{(s)}(y|x)$ denotes the conditional distribution of $Y$ given $X$ and state $s$.
By definition, $\rho(s_1, s_2|x) \leq 1$ for all $s_1$, $s_2$, and $x$. Additionally, under our model assumptions, there exists at least one input $x$ for any pair of states $s_1 \neq s_2$ such that $\rho(s_1, s_2|x) < 1$.
Extending this to an averaged Bhattacharyya coefficient over all possible inputs, we define
\begin{align*}
    \rho(s_1, s_2) := \frac{1}{|\Xc|} \sum_{x \in \Xc}\rho(s_1, s_2 | x),
\end{align*}
which satisfies $\rho(s_1, s_2) < 1$ as well for any $s_1 \neq s_2$.
The maximum averaged Bhattacharyya coefficient across all state pairs is then
\begin{align}
    \rho := \max_{s_1 \ne s_2} \rho(s_1, s_2), \label{eq:max_bhattacharya}
\end{align}
which by construction satisfies $\rho < 1$.

Also, the log-likelihood ratio (LLR) of $p^{(s)}$ relative to the base-state distribution is defined as
\begin{align}
	\Lambda^{(s)}(y|x) := \log\frac{p^{(s)}(y|x)}{p^{(0)}(y|x)}. \label{eq:def-LLR}
\end{align}
We then define the maximum LLR across all possible values of $y,x$ and $s$ as
\begin{align*}
	\gamma := \max_{x,y,s} \left| \Lambda^{(s)}(y|x) \right| < \infty
\end{align*}	
assuming the maximum is finite. 

Finally, we employ the notion of strong typicality in this work. For a random variable $X$, the strong typical set~\cite{Yeung2008} is defined as
\begin{align}
    \aep(X) = \left\{x^n: \sum_{x \in \Xc} \left|\pi(x|x^n)-p_X(x)\right|\le \e \right\}.\label{eq:strong_aep}
\end{align}
The set $\Pc_n(\Xc)$ denotes the collection of all types of $x^n$, and $\Tc_{p_X}^{(n)}$ with $p_X \in \Pc_n(\Xc)$ denotes the set of $x^n$ of type $p_X$.

\section{Main Results}

In this section, we present our main results and discuss their implications. The proof is given in the next section.

\begin{theorem}\label{thm:theorem1}
For a post-state set $\Sc$, a rate-delay pair $(R, \Delta)$ is achievable if for some $\{p_{X|S}(x|s), s\in\Sc\}$, 
\begin{align}
	R &< I(X_s;\Yt_s), \label{eq:thm1_rate}\\
	\Delta_{s} &< D(p^{(s)} \| p^{(0)} | p_{X_{s}}), \label{eq:thm1_delta}
\end{align}
for all $s\in\Sc$, where $(X_s, \Yt_s) \sim p(x|s)p(\yt|x, s)$.
\end{theorem}

We investigate the proposed rate-delay region in two extreme cases. First, we consider the case of pure communication in which $\Delta = 0$, i.e., there is no interest in delay performance. The pure communication problem alone introduces several unique challenges. To illustrate these, we compare our model to the compound channel with feedback studied in~\cite{wolfowitz1964, Shrader--Permuter2009}. This setting can be viewed as a special case of our model with a fixed change-point $\nu = 0$, an unknown transition state $s$, and $\Yt = Y$.
The main implication of the results in~\cite{wolfowitz1964, Shrader--Permuter2009} is that feedback can be used to estimate the channel state, allowing the transmitter to adapt its codewords to the estimated state. The resulting capacity of the compound channel with feedback is given by the minimum of state-dependent capacities, which is strictly larger than that of the compound channel without feedback. A similar phenomenon was observed for the case $Y \neq \Yt$ in~\cite{Chang--Wang--Erdogan--Bloch--2023} for a state-detection ISAC setting.

Our model extends these results to a setting where the state changes at an unknown time $\nu$. The following theorem characterizes the capacity in this regime, the proof of which is given in App.~\ref{app:pure_comm_proof}.

\begin{theorem}\label{thm:pure_comm}
For the case $\Delta = \mathbf{0}$, the channel capacity is
\begin{align}
C = \min_{s \in \Sc} \max_{p_{X_s}} I(X_s; \Yt_s). \label{eq:capacity_delta0}
\end{align}
\end{theorem}

Similar to the prior settings, the achievability is shown by using the echo channel observation to estimate the channel state, and then adapting the encoding to match the estimated state. For comparison, the compound channel without feedback (i.e., $Y = \emptyset$) attains
\begin{align}
R < \max_{p_X} \min_{s \in \Sc} I(X; \Yt_s). \label{eq:compound_rate}
\end{align}
By the minimax inequality,
\begin{align*}
\max_{p_X} \min_{s \in \Sc} I(X; \Yt_s) \le \min_{s \in \Sc} \max_{p_{X_s}} I(X_s; \Yt_s),
\end{align*}
which highlights the advantage of adaptive coding enabled by the echo channel in our setup.

In the second extreme case, we consider \( R = 0 \), corresponding to a pure QCD problem with control. This scenario has been previously studied in~\cite{Veeravalli--Fellouris--Moustakides2024}. For this case, we recover the results of~\cite[Thms.~1 and 2]{Veeravalli--Fellouris--Moustakides2024}, where the optimal asymptotic delay slope is characterized as
\begin{align*}
    \Delta_s &< \max_{P_X} D(p^{(s)}\| p^{(0)}|p_X) = \max_{x\in\Xc} D(p^{(s)}\| p^{(0)}|x),
\end{align*}
for \( s \in \Sc \), where the equality follows from the linearity of the conditional KL divergence in \( p_X \).

To elaborate further, when reliable message transmission is not required, the transmitter can fully exploit its control over the input symbols to optimize detection. If we specialize our achievable strategy for this case, it effectively alternates between pilot symbols and sensing symbols which are selected as the optimal detection symbols for the estimated state. Once the estimate converges after state transition, the sensing symbols can be chosen as the optimal statistics aligned with the true state, thereby recovering the optimal asymptotic delay slope derived in~\cite{Veeravalli--Fellouris--Moustakides2024}. This contrasts with the joint communication and detection scenario considered in this work, where the transmitted sequence must balance the dual objectives of reliable message delivery and timely change detection, thereby inducing a fundamental tradeoff.

Next, we specialize the result of Thm.~\ref{thm:theorem1} to the case when the communication channel is state independent, i.e., $p_{\Yt|X,S} = p_{\Yt|X}$. 
\begin{corollary} \label{cor:state_indep}
If the communication channel is state independent, then Thm.~\ref{thm:theorem1} implies that the set of rate-delay pairs $(R,\Delta)$ is achievable if for some $\{p(x|s): s\in\Sc\}$,
\begin{align}
R &< I(X_s;\Yt), \\
\Delta_s &< D(p^{(s)} \| p^{(0)} | p_{X_s}),  \label{eq:cor1}
\end{align}
for all $s\in\Sc$.
\end{corollary}
For comparison, we restate the achievable tradeoff region based on a open-loop strategy originally developed for state-independent communication channels given in~\cite{Seo--Lim2024}:
\begin{theorem}{\cite[Corollary 1]{Seo--Lim2024}}\label{thm:open-loop}
	The set of rate-delay pairs $(R, \Delta)$ is achievable if for some $p_X$,
	\begin{align}
		R &< I(X;Y), \label{eq:cor_rate}\\ 
		\Delta &< \min_{s \in \Sc } D(p^{(s)} \| p^{(0)} | p_X), \label{eq:cor_delta}
	\end{align}
	where $\Delta_s = \Delta$ for all $s\in\Sc$.
\end{theorem}

By comparing Cor.~\ref{cor:state_indep} and Thm.~\ref{thm:open-loop}, we observe that Thm.~\ref{thm:open-loop} imposes a universal bound on $\Delta$, meaning that the delay slope is uniformly constrained by the worst-case KL divergence across all states. Furthermore, the tradeoff between rate and delay in Thm.~\ref{thm:open-loop} relies on a single input distribution to simultaneously satisfy all mutual information and KL divergence constraints. In contrast, Cor.~\ref{cor:state_indep} enables state-dependent optimization of input distributions, resulting in a strictly larger achievable region. Section \ref{sec:BIBO_ex} demonstrates this gain for a binary channel example.

\begin{remark}
    The region in Thm.~\ref{thm:theorem1} can be extended to Gaussian channels with Gaussian input distributions using standard quantization techniques as described in~\cite{El-Gamal--Kim2011}. The specialized result to the Gaussian MIMO channels is given in Thm.~\ref{thm:mimo} with examples and discussions detailed in Sec.~\ref{sec:examples}.
\end{remark}

The next theorem states a lower bound (converse) on {\WADD} for the proposed JCCS strategy described in Sec.~\ref{sec:achievability}. The proof is given in App.~\ref{app:conv_proof}.
\begin{theorem} \label{thm:conv}
    For the proposed joint communication and QCD strategy generated by $\{p(x|s): s\in\Sc\}$, we have
    \begin{align*}
        \text{\sf WADD}(N,s) \ge \frac{ |\log \alpha| }{D(p^{(s)}\| p^{(0)}|p(x|s))}(1+o(1)),
    \end{align*}
for $s\in\Sc$.
\end{theorem}
The key implication of Thm.~\ref{thm:conv} is that it establishes the optimality of our detection strategy presented in Sec.~\ref{sec:achievability} under the proposed coding scheme. 
Notably, the delay slope derived in Thm.~\ref{thm:conv} matches the achievability result in Thm.~\ref{thm:theorem1}, confirming that the proposed detection algorithm is indeed asymptotically optimal for our codes.

\section{Achievability} \label{sec:achievability}

There are several parameters in the proof that require proper scaling with the block length $n$. 
These parameters will be rigorously defined when introduced; however, for ease of reference, we summarize the scaling laws of the relevant parameters in Tab.~\ref{tab:scaling}, where the parameters are assumed to grow unbounded as $k \to \infty$ while adhering to the specified scaling laws. A concrete example of parameter values satisfying these conditions is given by the following choices: $\eta  \propto t$, $L \propto t^3$, $k\propto t^5$, and $\alpha\propto \exp(-t^{5})$, where $t$ is a common scaling variable. 

\begin{table}[t]
\centering
\begin{tabular}{  c| c  }
  Variables  & Scaling laws \\ 
 \hline
 $\eta$ & $o(k)$, $\omega(\log(k))$\\
 $\eta L$ & $o(k)$,  $o(b)$\\
$b$ & $b=-\log(\alpha(L+1))$\\
$n$ & $n=(L+1)k$
\end{tabular}
\caption{Summary of scaling laws.}
\label{tab:scaling}
\end{table}

\subsection{Proposed joint communication and QCD strategy}
Our block code has a subblock structure, where a transmission signal of length $n$ consists of $k$ subblocks, each of length $L+1$. Let $\{\pi_q, q \in \Sc\}$ be a set of types on $\Xc^L$ indexed by $q \in \Sc$. Note that when $L$ is large enough, $\pi_q$ can approximate any $p_X$ on $\Xc$ within arbitrary accuracy. Recall that {$\Tc^{(L)}_{\pi_q}$ is the set of length-$L$ sequences with type $\pi_q$. For notational simplicity, a length $L$ sequence is represented as a vector, e.g., $\xv := (x_1, \ldots, x_L)$. We define a discrete memoryless channel (DMC) for an $L$-length vector as
$p(\tilde\yv|\xv, \sv) = \prod_{i=1}^L p_{\Yt|X}(\yt_i|x_i, s_i)$.

Fix $\{\pi_q, q \in \Sc \}$ and a pmf $p_Q(q)p_{\Xv|Q}(\xv|q)$ where $q \in \Sc$ and $p(\xv|q) = \textrm{Unif}\left( \Tc^{(L)}_{\pi_q} \right)$. 

\noindent \textbf{Codebook generation.} 
For each $q^k \in \Sc^k$, randomly and independently generate $2^{nR}$ sequences $\xv^k(m|q^k)$, $m \in [1:2^{nR}]$, each according to $\prod_{i=1}^k p_{\Xv|Q}(\xv_i|q_i)$, i.e., the codeword $\xv^k(m|q^k)$ consists of $k$ subblocks of length $L$.

We often omit $q^k$ and represent it by $\xv^k$ or $\xv^k(m)$ if the condition $q^k$ is clear from the context. 
We define the codebook $\Cc_{JCCS}^{(n)}=\{\xv_j(m|q): m\in[1:2^{nR}], q\in\Sc, j\in[1:k]\}$ as the JCCS codebook.

\noindent \textbf{Pilot symbols.} For each subblock $j \in [1:k]$, a pilot symbol $\breve{x}_j \in \Xc$ is generated uniformly and independently. These pilot symbols are common across all codewords, i.e., they do not depend on the message $m$.

\noindent \textbf{State estimation at transmitter.}  
Prior to transmitting symbols for the $j$-th subblock, the transmitter detects the state using maximum likelihood estimate (MLE) based on the channel outputs of the last $\eta$ pilot symbols, where $\eta$ is a positive integer. That is, for subblock indices $j \in [\eta+1:k]$, the MLE is computed as:
\begin{align}
	\sh_j = \argmax_{s \in \mathcal{S} } \sum_{\ell=j-\eta}^{j-1} \log p^{(s)}_{Y|X}(\breve{y}_\ell|\breve{x}_\ell), \label{eq:MLE}
\end{align}
where $\breve{x}_j$ and $\breve{y}_j$ are the pilot and its corresponding channel output, respectively. For indices $j \in [1:\eta]$, the state estimate $\sh_j$ is selected uniformly at random from $\Sc$. By construction, the state estimate $\sh_j$ depends only on the $\eta$-previous pilot symbols and is conditionally independent of the transmitted codeword, given these pilot symbols and observations.

\noindent \textbf{Encoding at transmitter.} 
To send $m \in [1:2^{nR}]$, the transmitter alternates between sending codeword subblocks and pilot symbols. For each subblock $j \in [1:k]$, the transmitter first sends the codeword $\xv_j(m|\sh_j)$, where $\sh_j$ is the estimated state for subblock $j$, determined using the state estimation procedure. This is immediately followed by the transmission of the random pilot symbol $\breve{x}_j$. Consequently, the overall transmission signal $x^n$ is of length $n=k(L+1)$ and alternates between codeword blocks and pilot symbols as 
$x^n = (\xv_1, \breve{x}_1, \xv_2, \breve{x}_2, \ldots, \xv_k, \breve{x}_k)$.

\begin{remark}
When referring to a codebook $\Cc_{JCCS}^{(n)}$ as fixed, we mean that the codewords $\xv_j(m|q)$ have been sampled for all $m \in [1:2^{nR}]$, $j \in [1:k]$, and $q \in \Sc$. However, it is important to note that the \emph{transmitted} codeword is random, even with a fixed codebook. This randomness arises because the transmission is based on $\xv_j(m|\Sh_j)$, where $\Sh_j$ is a random variable determined through the state estimation process.
\end{remark}

\noindent\textbf{Sending state estimation sequence to receiver.}  
After completing the transmission, the transmitter sends the state estimation sequence $\sh^k$ to the receiver using either a universal source coding scheme or a fixed-length source code, such as those established in~\cite{Han--Verdu1993, Han2003}. Since the number of possible states is finite and the estimated sequence $\sh^k$ is highly structured, its spectral sup-entropy rate tends to zero (see App.~\ref{app:entropy_rate}). Consequently, the required rate for transmitting $\sh^k$ becomes asymptotically negligible.

Specifically, the asymptotic compression rate converges to the entropy rate of the source. Furthermore, the entropy rate of the state estimation sequence diminishes as $k \to \infty$, i.e., 
\begin{align}
\lim_{k\to\infty}\frac{1}{k}H(\Sh^k)=0, \label{eq:ER1}
\end{align}
which implies that the compression rate for sending $\sh^k$ becomes negligible in the limit. 
As the source rate is negligible, the required data rate for reliable communication of the estimated state sequence is also negligible. 
The proof of \eqref{eq:ER1} is in App.~\ref{app:entropy_rate}.

\noindent\textbf{Message decoding at receiver.} The decoder declares that $\mh\in[1:2^{nR}]$ was sent if it is the unique message such that 
\begin{align}
	(\sh^k, \xv^k(m|\sh^k), {\tilde\yv}^k) \in \aepk(Q, \Xv, \tilde\Yv),\label{eq:joint_typical_decoding}
\end{align}
otherwise, it declares an error. Here, $\sh^k$ is the sequence of ML estimates sent by the transmitter. The typical set $\aepk(Q, \Xv, \tilde\Yv)$ is defined with respect to the joint distribution $p_Q(q)p(\xv|q)p(\tilde\yv|\xv, q)$ 
where $p_Q$ is deterministic and satisfies $p_Q(\sh_k)=1$, i.e., assigning probability one to the last ML estimate. We emphasize that $\sh^k$ is not an i.i.d.~sequence with respect to $p_Q$. As a result, the standard decoding analysis, which typically assumes alignment between the test sequence and the typical set distribution, does not apply directly and requires additional proof steps. We also note that the pilots and the pilot-induced outputs $\breve{y}^k$ are indirectly included in the decoding step via $\sh^k$.

\noindent\textbf{Quickest change detection.}
For QCD, we introduce a variation of the cumulative sum (CuSum) algorithm, tailored for our coding strategy. 
This modified algorithm, referred to as the Subblock CuSum (\SCS) algorithm, updates the CuSum statistic only at the end of each subblock $j \in [1:k]$. 
The {\SCS} statistic is as follows:
\begin{align*}
	\Wv_{j, \xv^k} = 
	\begin{cases} 
		0, & \text{if } j \le \eta, \\ 
		\max\{\Wv_{j-1, \xv^k}, 0\} + \Lambda^{(\Sh_j)}(\Yv_j|\xv_j), & \text{if } j > \eta,
	\end{cases}
\end{align*}
where $\Sh_j$ is the state estimate of subblock $j$. While the definition may appear to depend only on the current input-observation pairs $(\xv_j, \Yv_j)$, the statistic $\Wv_{j, \xv^k}$ for $j > \eta$ is also influenced by the past pilots through the state estimate $\Sh_j$.

The stopping time, measured in terms of the subblock indices, $\Nv_{b, \xv^K}$, is defined as 
\begin{align}
	\Nv_{b, \xv^k} := \inf \{ j > \eta : \Wv_{j, \xv^k} \ge b \},\label{eq:block_stoptime}
\end{align}
where $b$ is a predefined threshold. By definition, it is obvious that $\Nv_{b, \xv^k} > \eta$.
Since, the delay time metric of interest is based on symbol time, the subblock index delay is translated  to symbol-wise delay $N_{b,\xv^k}$ as
\begin{align}
    N_{b,\xv^k} = \Nv_{b, \xv^k}(L+1).\label{eq:sym_stoptime}
\end{align}
where $L+1$ is the length of each subblock, including the pilot symbol. 
Specifically, since $\Nv_{b, \xv^k}$ represents the block index $j$ of the earliest subblock crossing the threshold $b$, the stopping time in symbols is conservatively declared as $\Nv_{b, \xv^k}(L+1)$.

Before presenting the performance analysis of the proposed strategy, we state a lemma that provides an upper bound on the probability of state estimation error. This bound will be used frequently in the achievability proof. The detailed proof of this lemma is deferred to App.~\ref{app: state_est_error_bound}.

\begin{lemma} \label{lem:MLE_error_unconditional}
Suppose that the true state is $s^\star$, the change point occurs at $\nu$, and $\nu$ is contained in the $j_\nu$-th subblock. Further, assume that the maximum likelihood estimate (MLE) in \eqref{eq:MLE} is based on $\eta$ observations after $\nu$. Then, for any $j \geq j_\nu + \eta$, the probability of estimation error for $\sh_{j}$ satisfies
	\begin{align*}
		P_{\nu}^{(s^\star)} ( \Sh_j \ne s^\star ) \le (|\Sc|-1) \rho^{\eta},
	\end{align*}
where $\rho$ is the maximum averaged Bhattacharyya coefficient defined in~\eqref{eq:max_bhattacharya}. 
\end{lemma}
Since $\rho < 1$, the probability of error tends to zero at least exponentially fast as $\eta \to \infty$. 

\subsection{Communication probability of error analysis}
Recall that the decoder recovers a transmitted message using joint typicality, as defined in \eqref{eq:joint_typical_decoding}. Without loss of generality, suppose that $m=1$ was sent and that the true post-transition state after $\nu$ is $s^\star \in \Sc$. Let $I_j = \mathbf{1}\{\Sh_j=s^\star\}$ be the indicator random variable for the correct MLE and 
\begin{align*}
	\Ec_1 &\!=\! \{(\Sh^k, \Xv^k(1), \tilde\Yv^k) \not\in \aepk (Q, \Xv, \Yv) \}\\
	\Ec_2 &\!=\! \{(\Sh^k, \Xv^k(m), \tilde\Yv^k) \!\in\! \aepk (Q, \Xv, \Yv) \text{ for some } m\neq 1\}.
\end{align*}
Then, the probability of decoding error $\P(\Ec)$, where $\Ec = \Ec_1 \cup \Ec_2$, can be bounded as follows.
\begin{align*}
	\P(\Ec) &= \P(I_k=1) \P(\Ec|I_k=1) + \P(I_k=0) \P(\Ec|I_k=0) \\
	&\le \P(\Ec|I_k=1) + \P(I_k=0) \\
	&\le \P(\Ec|I_k=1) + (|\Sc|-1) \rho^{\eta} ~~~ \text{by Lem.~\ref{lem:MLE_error_unconditional}} \\
	&\le \P(\Ec_1|I_k=1) + \P(\Ec_2|I_k=1) + \e_k.
\end{align*}
In standard random coding proofs, the test sequences are drawn from i.i.d. distributions that define the typical set, and thus, $\P(\Ec_1)$ vanishes as $k\to\infty$ simply due to the law of large numbers. However, the analysis in our case is more intricate because $\sh^k$ consists of MLEs derived from feedback-induced observations. Consequently, $\sh^k$ is not i.i.d. from $p_Q$, and the law of large numbers cannot be simply applied. To progress in our case, consider the underlying probability measure  $P_{\nu}^{(s^\star)} (\cdot)$ which corresponds to the distribution under the assumption of change state $s^\star$ at time $\nu$. The probability of the first error event can be bounded as follows, where the condition $I_k=1$ is omitted for simplicity.
\begin{align*}
	&\P (\Ec_1|I_k=1) = P_{\nu}^{(s^\star)} \left( (\Sh^k, \Xv^k(1), \tilde\Yv^k) \not\in \aepk (Q, \Xv, \Yv) \right) \\
	&= P_{\nu}^{(s^\star)} \left( \Sh^k \notin \aepk(Q) \right)\\
        &+ P_{\nu}^{(s^\star)} \left( (\Sh^k, \Xv^k(1), \tilde\Yv^k) \not\in \aepk (Q, \Xv, \Yv) \Big| \Sh^k \in \aepk(Q) \right) \\
	&\stackrel{(a)}{\le} P_{\nu}^{(s^\star)} \left( \Sh^k\notin \aepk(Q) \right) + \e_k \\
	&\stackrel{(b)}{=} P_{\nu}^{(s^\star)} \Bigg( \sum_{q \in \Sc} \left| \pi(q|\Sh^k) - p_Q(q) \right| > \e \Bigg) + \e_k,\\
    &\stackrel{(c)}{=} P_{\nu}^{(s^\star)} \Big( \left( 1-\pi(s^\star|\Sh^k) \right) + \sum_{q \ne s^\star} \pi(q|\Sh^k) > \e \Big) + \e_k \\
	&\le P_{\nu}^{(s^\star)} \left( 1-\pi(s^\star|\Sh^k) > \frac{\e}{2} \right) + P_{\nu}^{(s^\star)} \Big( \sum_{q \ne s^\star} \pi(q|\Sh^k) > \frac{\e}{2} \Big) \\
    &\quad + \e_k \\
    &\stackrel{(d)}{=} 2 P_{\nu}^{(s^\star)} \left( 1-\pi(s^\star|\Sh^k) > \frac{\e}{2} \right) + \e_k,
\end{align*}
where $(a)$ is due to the conditional typicality lemma~\cite{El-Gamal--Kim2011} adapted to the strong typicality, $(b)$ is by the definition of the strong typicality, $(c)$ is due to the fact that $p_Q(q)$ is a deterministic distribution such that $p_Q(\sh_k)=p_Q(s^\star)=1$ conditioned on $I_k=1$, and $(d)$ follows since the two probabilities are indeed the same.
Then, since $\nu$ is contained in the $j_\nu$-th subblock, 
\begin{align*}
	&P_{\nu}^{(s^\star)} \left( 1-\pi(s^\star|\Sh^k) > \frac{\e}{2} \right)\\
    &= P_{\nu}^{(s^\star)} \Big( 1- \frac{1}{k} \sum_{j=1}^k \mathbf{1}\{ \Sh_j = s^\star \} > \frac{\e}{2} \Big) \\
	&= P_{\nu}^{(s^\star)} \Bigg( \frac{1}{k} \sum_{j=1}^k \mathbf{1}\{ \Sh_j \ne s^\star \} > \frac{\e}{2} \Bigg) \\
	&\stackrel{(e)}{\le} \frac{ \E_{\nu}^{(s^\star)} \left[ \frac{1}{k} \sum_{j=1}^k \mathbf{1}\{ \Sh_j \ne s^\star \} \right] }{\e/2} \le \frac{ \frac{1}{k} \sum_{j=1}^k P_{\nu}^{(s^\star)} ( \Sh_j \ne s^\star ) }{\e/2} \\
	&\stackrel{(f)}{\le} \frac{2}{k} \cdot \frac{ j_\nu + \eta - 1 + k (|\mathcal{S}|-1) \rho^\eta }{\e} \to 0 ~~ \text{if } \eta = o(k),
\end{align*}
where $(e)$ is due to the Markov inequality, and $(f)$ follows from simply assuming the first $(j_\nu+\eta-1)$ MLEs are all errors and Lem.~\ref{lem:MLE_error_unconditional}.

Next, we bound the probability $\P(\Ec_2|I_k=1)$ as follows, where the condition $I_k=1$ is omitted.
\begin{align*}
	&\P(\Ec_2|I_k=1) \le \sum_{m=2}^{2^{nR}} P_{\nu}^{(s^\star)} \left( (\Sh^k, \Xv^k(m), \tilde\Yv^k) \in \aepk \right) \\
	&=\sum_{m=2}^{2^{nR}}\sum_{(q^k, \tilde\yv^k) \in \aepk(Q, \tilde\Yv)} P_{\nu}^{(s^\star)} \left( \Sh^k=q^k, \tilde\Yv^k=\tilde \yv^k \right) \\ 
    &\quad \times  \sum_{\xv^k\in\aepk(\Xv|q^k,\tilde \yv^k)}p(\xv^k|q^k) \\
	&\le \sum_{m=2}^{2^{nR}} \sum_{(q^k, \tilde\yv^k) \in \aepk(Q, \tilde\Yv)} P_{\nu}^{(s^\star)} \left( \Sh^k=q^k, \tilde\Yv^k=\tilde \yv^k \right) \\ &\quad \times \sum_{\xv^k\in\aepk(\Xv|q^k,\tilde \yv^k)}2^{-k(H(\Xv|Q)-\d(\e))} \\
	&\le \sum_{m=2}^{2^{nR}} \sum_{(q^k, \tilde\yv^k) \in \aepk(Q, \tilde\Yv)} P_{\nu}^{(s^\star)} \left( \Sh^k=q^k, \tilde\Yv^k=\tilde \yv^k \right) \\ 
    &\quad \times  2^{k(H(\Xv|Q,\tilde \Yv)+\d(\e))}2^{-k(H(\Xv|Q)-\d(\e))}\\
	&\le 2^{nR} \cdot 2^{-k(I(\Xv;\tilde\Yv|Q)-2\d(\e))}.
\end{align*}
Thus, $\P(\Ec_2 | I_k=1)$ tends to zero as $k \to \infty$ if
\begin{align*}
	R&<\frac{1}{L+1}I(\Xv;\tilde\Yv|Q)=\frac{1}{L+1}I(\Xv;\tilde\Yv|Q=s^\star).
\end{align*}
Finally, since
$p(\xv, \tilde{\yv}|Q=s^\star) = \prod_{i=1}^L p(x_i|s^\star)p(\yt_i|x_i, s^\star)$
by Thm.~7 in~\cite{Tandon--Motani--Varshney2016}, 
\begin{align*}
	\lim_{L\to\infty}\frac{1}{L}I(\Xv;\tilde\Yv|Q=s^\star) = I(X_{s^\star};\Yt_{s^\star}),
\end{align*}
where $(X_{s^\star},\Yt_{s^\star})\sim p(x|s^{\star})p(\yt|x, s^\star)$.

Overall, the probability of decoding error tends to zero as $n\to\infty$ if
$R<I(X_{s^\star};\Yt_{s^\star})$ for some $X_{s^\star}\sim p(x|s^\star)$. 
However, since the rate $R$ is determined without prior knowledge of $s^\star$ and $\nu$, we have to adopt a conservative approach and set the rate to account for the worst state realization:
\begin{align*}
	R &< \min_{s} \max_{p_{X_s}} I(X_s;\Yt_{s})
\end{align*}
which is achievable. It is important to emphasize that setting the rate $R$ in advance does not imply the absence of adaptive coding. For a fixed rate $R$, the encoder transmits subblocks based on the state-estimates, which enables the ``$\min\max$'' structure in the rate constraint. This structure is superior to the ``$\max\min$'' form that arises when non-adaptive codes are employed.

\subsection{QCD analysis}

We begin the proof with the following lemma that bounds the false alarm rate.
\begin{lemma} \label{lem:FAR}
	Consider a sequence of JCCS strategies $\{\Cc^{(n)}_{JCCS}\}_n$. Then, by taking $b = - \log (\alpha (L+1)) = |\log (\alpha (L+1))|$, the false alarm rate is bounded by
	\begin{align*}
		\emph{\FAR}(N_b) =  \limsup_{n \to \infty} \max_{m\in[1:2^{nR}]} \frac{1}{\E_{\infty}[ N_{b}(m) ]} \le \alpha.
	\end{align*}
\end{lemma}
\begin{IEEEproof}
Note that our CuSum statistic can be expressed in an exponential form as
\begin{align*}
    \exp (\Wv_{j}) &= \begin{cases}
        1 & \text{if } j \le \eta \\
            \max \left\{ e^{\Wv_{j-1}}, 1 \right\} \frac{ p_{\Yv|\Xv}^{(\Sh_j)}(\Yv_j | \Xv_j(m|\Sh_j)) }{ p_{\Yv|\Xv}^{(0)}(\Yv_j|\Xv_j(m|\Sh_j))} & \text{if } j > \eta.
    \end{cases}
\end{align*}
Next, we define a new statistic that modifies the Shiryaev-Roberts statistic: For $j > \eta$, 
\begin{align*}
    \Rv_{j} = \left( \Rv_{(j-1)} + 1 \right) \frac{ p_{\Yv|\Xv}^{(\Sh_j)}(\Yv_j | \Xv_j(m|\Sh_j)) }{ p_{\Yv|\Xv}^{(0)}(\Yv_j|\Xv_j(m|\Sh_j))},
\end{align*}
where $\Rv_{\eta} = 1$. Note that when $k \to \infty$, the sequence $\{\Rv_{j} - j: j > \eta \}$ forms a martingale under $P_{\infty}$ with respect to $\mathcal{F}_{j}(m)$. 
Then, for $k \to \infty$, by the optional stopping theorem~\cite{Durrett2019},
$\E_{\infty} \left[ \Rv_{\Nv_b} - \Nv_{b} \right] = \Rv_{\eta} - \eta = 1-\eta$.
Using the fact that $\Rv_{j} \ge \exp (\Wv_{j})$ and by~\eqref{eq:sym_stoptime},
\begin{align*}
    \E_{\infty} [ N_{b} ] &= (L+1) \E_{\infty} [ \Nv_{b} ] \\
    &= (L+1) ( \E_{\infty} \left[ \Rv_{\Nv_{b}} \right] + \eta - 1 ) \\
    &\ge (L+1) ( \E_{\infty} \left[ \exp (\Wv_{\Nv_{b}}) \right] + \eta - 1 ) \\
    &\ge (L+1) \E_{\infty} \left[ \exp (\Wv_{\Nv_{b}}) \right] \\
    &\ge (L+1) \exp (b).
\end{align*}
Note that $N_b$, $\Nv_b$, $\Rv_{\Nv_b}$, $\Wv_{\Nv_b}$ are all functions of the message $m$ which is omitted for  simplicity. Nonetheless, the final bound is independent of the message, providing a universal bound. Thus, 
\begin{align*}
    \FAR(N_b) &\le \exp(-b - \log (L+1))=\alpha
\end{align*} 
where $b = -\log (\alpha (L+1))$.
\end{IEEEproof}

Next, we have the bound on $\WADD$ as follows.
\begin{theorem}
For the JSCC strategy, $\WADD$ is bounded by
	\begin{align*}
	&\limsup_{n \to \infty} \max_{m\in[1:2^{nR}]} \esssupe \E_{\nu, \xv^k}^{(s)}[ (N_{b, \xv^k}-\nu+1)^+ | \Fc_{\nu-1} ] \\
        &\quad \le \frac{b}{D( p^{(s^\star)} \| p^{(0)} | p_{X|s^\star} )} (1+o(1)).
	\end{align*}
\end{theorem}
\begin{IEEEproof}
	The following inequalities hold:
	\begin{align*}
		&\limsup_{n \to \infty} \max_{m\in[1:2^{nR}]} \esssup \E_{\nu, \xv^k}^{(s)}[ (N_{b, \xv^k}-\nu+1)^+ | \Fc_{\nu-1} ] \\
		&\le \sup_{ x^w } \E_{1, \xv^k}^{(s)} \left[ N_{b, \xv^k} |X^w = x^w \right] ~~~ \text{by Lem.~\ref{lem:WADD_UB}} \\
		&\le \frac{b}{D( p^{(s)} \| p^{(0)} | p_{X|s} )} ( 1+o(1) ) ~~~ \text{by Thm.~\ref{thm:Nb_bound}}
	\end{align*}
\end{IEEEproof}

The remaining of this section is devoted to proving 
Lem.~\ref{lem:WADD_UB} and Thm.~\ref{thm:Nb_bound}.

\begin{lemma} \label{lem:WADD_UB}
For any change point $\nu$, sequence of JCCS codebooks $\Cc_{JCCS}^{(n)}$, and some large enough threshold $b > 0$,
\begin{align*}
  &\limsup_{n \to \infty} \max_{m\in[1:2^{nR}]} \esssupe \E_{\nu, \xv^k}^{(s)}[ (N_{b, \xv^k}-\nu+1)^+ | \Fc_{\nu-1}(m) ]\\
  &\qquad\le \limsup_{n \to \infty}  \sup_{ x^w } \E_{1, \xv^k}^{(s)} \left[ N_{b, \xv^k} |X^w = x^w \right]
\end{align*}
where $w=\eta(L+1)$.
\end{lemma}

\begin{IEEEproof}
To facilitate the proof, we introduce a hypothetical $\nu$-aware subblock CuSum statistic $\Wv_{j}^{\nu}$ and a stopping rule $N_{b}^{\nu}$ (or $\Nv_b^{\nu}$ for subblock form) as analytical tools. The 
$\nu$-aware CuSum statistic $\Wv_{j}^{\nu}$ is updated every subblock by 
\begin{align}
    &\Wv_{j}^{\nu} \nn \\
    &= \begin{cases}
        0 & \text{ if } j \le j_\nu + \eta-1, \\
        \max\{ \Wv_{j-1}^{\nu}, 0 \} + \Lambda^{(\Sh_j)}(\Yv_j|\xv_j) & \text{ if } j > j_\nu + \eta-1, 
    \end{cases} \label{eq:W_nu}
\end{align}
where a nontrivial update starts from the $(j_\nu + \eta)$-th subblock.
The corresponding stopping rule in terms of the subblock indices, $\Nv^{\nu}_{b}$, is defined analogously to $\Nv_b$ in \eqref{eq:block_stoptime} as:
\begin{align}
	\Nv^{\nu}_{b} := \inf \{ j > j_\nu+\eta-1 : \Wv^{\nu}_{j} \ge b \}.\label{eq:nu_aware_block_stoptime}
\end{align}
We translate the block index delay to symbol-wise $\nu$-aware delay $N^\nu_{b}$ akin to $N_b$ in \eqref{eq:sym_stoptime}, as $N^{\nu}_{b} = \Nv^{\nu}_{b}(L+1).$
Notably, since non-trivial updates for $\Wv^\nu_j$ begin $(j_\nu-1)$ blocks later than that for $\Wv_j$, the relation $N_{b} \le N_{b}^{\nu}$ holds.
Then, for any $\nu \ge 1$, post state $s \in \Sc$, and $m \in[1:2^{nR}]$, we have
\begin{align*}
	&\E_{\nu}^{(s)}[ (N_{b}-\nu+1)^+ | \mathcal{F}_{\nu-1}(m) ]\\
    &\stackrel{(a)}{\le} w + \E_{\nu}^{(s)}[ (N_{b}-\nu+1-w)^+ | \mathcal{F}_{\nu-1}(m) ] \\
	&\stackrel{(b)}{\le} w + \E_{\nu}^{(s)}[ N_{b}^{\nu}-\nu+1-w | \mathcal{F}_{\nu-1}(m) ] \\
    &\stackrel{(c)}{\le} w + \E_{\nu}^{(s)}[ N_{b}^{\nu}-\nu+1-w | \mathcal{F}_{\nu-1}(m) ]
\end{align*}
where $(a)$ follows since $z \le w + (z-w)^+$ for any $z,w \ge 0$ and $(b)$ follows from two facts that $N_{b} \le N_{b}^{\nu}$ and $N_{b}^{\nu} \ge \nu + w - 1$ by definition. To see $(c)$, note that the statistic $\Wv_j^\nu$ in \eqref{eq:W_nu} is dependent only on observations induced by $\xv_j, j > j_\nu + \eta-1$, where the composition of $\xv_j$ is determined by $\Sh_j$ that is dependent only on pilots after $\nu-1$. Since pilots are i.i.d. and shared across codewords, the expression is independent of $m$. Hereafter, we omit the dependency on $m$ and write $\Fc_{\nu-1} = \Fc_{\nu-1}(m)$. Then,
\begin{align*}
	&\E_{\nu}^{(s)}[ (N_{b}-\nu+1)^+ | \mathcal{F}_{\nu-1}(m) ]\\
    &\le w + \E_{\nu}^{(s)}[ N_{b}^{\nu}-\nu+1-w | \mathcal{F}_{\nu-1} ] \\
    &= \E_{\nu}^{(s)}[ N_{b}^{\nu}-\nu+1 | \mathcal{F}_{\nu-1} ] \\
    &\stackrel{(d)}{=} \E_{\nu}^{(s)} \left[ \E_{\nu}^{(s)} \left[ N_{b}^{\nu}-\nu+1 | X_{-w}^{(j_\nu+\eta-1)(L+1)}, \mathcal{F}_{\nu-1} \right] \Big| \mathcal{F}_{\nu-1} \right] \\
	&\stackrel{(e)}{=} \E_{\nu}^{(s)} \left[ \E_{\nu}^{(s)} \left[ N_{b}^{\nu}-\nu+1 | X_{-w}^{(j_\nu+\eta-1)(L+1)} \right] \Big| \mathcal{F}_{\nu-1} \right] \\
	&\stackrel{(f)}{\le} \sup_{ x^w } \E_{\nu}^{(s)} \left[ N_{b}^{\nu} - \nu+1 |X_{-w}^{(j_\nu+\eta-1)(L+1)} = x^w \right] \\
	&\stackrel{(g)}{\le} \sup_{ x^{w} } \E_{1}^{(s)} \left[ N_{b} |X^w = x^w \right],
\end{align*}
where $X_{-w}^{(j_\nu+\eta-1)(L+1)} := X_{(j_\nu-1)(L+1)+1}^{(j_\nu+\eta-1)(L+1)}$ is the subsequence of length $w = \eta(L+1)$ ending with the last symbol $X_{(j_\nu+\eta-1)(L+1)}$. Step $(d)$ follows from the law of iterated expectation, $(e)$ follows since $N_b^\nu$ is a function of $\Nv_b^\nu$ and that conditioned on $X_{-w}^{(j_\nu+\eta-1)(L+1)}$, $\Nv_b^{\nu}$ is independent of $\Fc_{\nu-1}$ since it only depends on observations in the subblocks $j > j_\nu + \eta-1$, and also given $X_{-w}^{(j_\nu+\eta-1)(L+1)}$, the MLEs $\Sh_j$, $j \le j_\nu + \eta$ are also independent of $\Nv_b^{\nu}$. To be more precise, the conditional independence of $\Nv_b^\nu$ and $\Sh_j$, $j \le j_\nu + \eta$ is due to the fact that in the subsequence $X_{-w}^{(j_\nu+\eta-1)(L+1)}$, there are at least $\eta$ pilot symbols that completely characterize the estimate $\Sh_j$. 
Step $(f)$ follows since replacing a subsequence $X_{-w}^{(j_\nu+\eta-1)(L+1)}$ of the whole transmission sequence $X^n$ with the worst subsequence (which also includes pilot symbols; hence, also replacing pilots with worst pilots) can only increase the delay. 

Finally, step $(g)$ is due to the following. We first show the relation
\begin{align}
    &\E_{\nu}^{(s)} \left[ \Nv_{b}^{\nu} - j_\nu+1 |X_{-w}^{(j_\nu+\eta-1)(L+1)} = x^w \right] \nn \\
    &= \E_{1}^{(s)} \left[ \Nv_{b}^{\nu=1} |X^w = x^w \right]. \label{eq:shift_equalization}
\end{align}
To show this, we compare the statistics $\Wv^\nu_j$ for some $\nu>1$ with $\Wv^{\nu=1}_j$, i.e., the $\nu$-aware statistic when $\nu=1$. First, note that both statistics have a trivial update stage, i.e., $\Wv_j^{\nu}=0$ for $j\le j_\nu+\eta-1$. For some change point $\nu \ge 1$, the subsequence $X_{-w}^{(j_\nu+\eta-1)(L+1)}$ is the last $w$ symbols ending exactly at the end of the $(j_\nu + \eta-1)$-th subblock.
The statistic $\Wv_j^{\nu}$ starts a nontrivial update from subblock $j_\nu+\eta$, i.e., the next subblock. 
Similarly, for the case $\Wv_j^{\nu=1}$, a nontrivial update starts from subblock $\eta+1$. 
Since $\Wv_{j_\nu+\eta}^{\nu}$ for some $\nu>1$ and $\Wv_{ \eta+1}^{\nu=1}$ both initialize with a trivial update and transition to the first nontrivial update based on the closest past $\eta$ pilots, we have that the distribution of $\Wv_{j}^{\nu}$, $j\ge j_\nu+\eta-1$ given $X_{-w}^{(j_\nu+\eta-1)(L+1)}=x^w$ and $\Wv_{j}^{\nu=1}$, $j\ge 1$ given $X^w=x^w$ are the same.
Since the relative delay for both cases are based on the same statistics, we have the relation in~\eqref{eq:shift_equalization}. Then,
\begin{align*}
    &\E_{\nu}^{(s)} \left[ N_{b}^{\nu} - \nu+1 |X_{-w}^{(j_\nu+\eta-1)(L+1)} = x^w \right]\\
    &\le \E_{\nu}^{(s)} \left[ (L+1)(\Nv_{b}^{\nu} - j_\nu+1) |X_{-w}^{(j_\nu+\eta-1)(L+1)} = x^w \right]\\
    &= \E_{\nu}^{(s)} \left[ (L+1)\Nv_{b}^{1} |X^w = x^w \right]\\
    &= \E_{\nu}^{(s)} \left[ N_{b}^{1} |X^w = x^w \right],
\end{align*}
where the first inequality is due to
\begin{align*}
    N_b^\nu-\nu+1 &= (L+1)\left(\Nv_b^\nu-\frac{\nu-1}{L+1}\right) \\
    &\le (L+1)(\Nv_b^\nu-j_\nu+1),
\end{align*}
\end{IEEEproof}

\begin{lemma} \label{lem:MLE_error_conditional}
	For any codeword $\xv^k$, if pilot symbol $\breve{X}_j \in \mathcal{X}$ is chosen uniformly at random for $j \in [1:k]$, then for all $j > 2\eta$,
	$$\P_{1, \xv^k}^{(s)} ( \Sh_j \ne s | \Fc_{j(L+1)-w-1}, x^w ) \le (|\Sc|-1) \rho^\eta.$$
\end{lemma}
\begin{IEEEproof}
The proof follows directly from Lem.~\ref{lem:MLE_error_unconditional} and 
\begin{align*}
	\P_{1, \xv^k}^{(s)} ( \Sh_j \ne s | \Fc_{j(L+1)-w-1}, x^w )= \P_{1, \xv^k}^{(s)} ( \Sh_j \ne s ) 
\end{align*}
for $j>2\eta$.
\end{IEEEproof}
A simple interpretation of the above is that since there are $\eta$ pilot-induced observations in the past $w$ window, they contribute to the $\rho^\eta$ factor of the upper bound.

\begin{lemma} \label{lem:drift_LB}
	Suppose that the true post state is $s^\star$ and $\nu=1$. Then, the expected drift of the log-likelihood ratio for the subblock $j > 2\eta$ is
	\begin{align*}
		&\E_{1}^{(s^\star)} \left[ \Lambda^{(\Sh_j)}(\Yv_j|\xv_j)  \big| \Fc_{j(L+1)-w-1}, x^w \right] \\
		&\ge L D( p^{(s^\star)} \| p^{(0)} |  p_{X|s^\star} ) - L \gamma (|\mathcal{S}|-1) \rho^\eta.
	\end{align*}
	Moreover, for any $\epsilon > 0$, there exist a large enough $L$ and $\eta$ such that
	\begin{align*}
		&\frac{1}{L+1} \E_{1}^{(s^\star)} \left[ \Lambda^{(\Sh_j)}(\Yv_j|\xv_j) | \Fc_{j(L+1)-w-1}, x^w \right]\\
        &\ge D( p^{(s^\star)} \| p^{(0)} | p_{X|s^\star} ) - \epsilon.
	\end{align*}
\end{lemma}
\begin{IEEEproof}	
	Suppose that $\xv_j$ of composition $p_{X|\sh_j}$ was sent in the $j$-th subblock. To emphasize its composition, we write $\xv_j = \xv_j^{(\sh_j)}$. Then,
{\small
\begin{align*}
	&\E_{1}^{(s^\star)} \left[ \Lambda^{(\Sh_j)}(\Yv_j|\xv_j^{(\Sh_j)}) | \Fc_{j(L+1)-w-1}, x^w \right] \\
		&= \sum_{s} \P_{1}^{(s^\star)} ( \Sh_j = s | \Fc_{j(L+1)-w-1}, x^w )\\
        &~~~~ \times\E_{1}^{(s^\star)} \left[ \Lambda^{(s)}(\Yv_j|\xv_j^{(s)}) | \Sh_j = s, \Fc_{j(L+1)-w-1}, x^w \right] \\
		&\stackrel{(a)}{=} \sum_{s} \P_{1}^{(s^\star)} ( \Sh_j = s | \Fc_{j(L+1)-w-1}, x^w ) \E_{1}^{(s^\star)} \left[ \Lambda^{(s)}(\Yv_j|\xv_j^{(s)}) \right] \\
		&= \P_{1}^{(s^\star)} ( \Sh_j = s^\star | \Fc_{j(L+1)-w-1}, x^w ) \E_{1}^{(s^\star)} \left[\Lambda^{(s^\star)}(\Yv_j|\xv_j^{(s^\star)}) \right] \\
		& + \sum_{s \ne s^\star} \P_{1}^{(s^\star)} ( \Sh_j = s | \Fc_{j(L+1)-w-1}, x^w ) \E_{1}^{(s^\star)} \left[ \Lambda^{(s)}(\Yv_j|\xv_j^{(s)}) \right] \\
		&\ge \P_{1}^{(s^\star)} ( \Sh_j = s^\star | \Fc_{j(L+1)-w-1}, x^w ) \E_{1}^{(s^\star)} \left[\Lambda^{(s^\star)}(\Yv_j|\xv_j^{(s^\star)})\right] \\
		& + \sum_{s \ne s^\star} \P_{1}^{(s^\star)} ( \Sh_j \ne s^\star | \Fc_{j(L+1)-w-1}, x^w ) \\
        &\times \min_{s \ne s^\star}\E_{1}^{(s^\star)} \left[ \Lambda^{(s)}(\Yv_j|\xv_j^{(s)})\right] \\
		&= \E_{1}^{(s^\star)} \left[ \Lambda^{(s^\star)}(\Yv_j|\xv_j^{(s^\star)})\right] - \P_{1}^{(s^\star)} ( \Sh_j \ne s^\star | \Fc_{j(L+1)-w-1}, x^w ) \\
        &~~~\times\left( \E_{1}^{(s^\star)} \left[\Lambda^{(s^\star)}(\Yv_j|\xv_j^{(s^\star)})\right] - \min_{s \ne s^\star} \E_{1}^{(s^\star)} \left[ \Lambda^{(s)}(\Yv_j|\xv_j^{(s)}) \right] \right) \\
		&\stackrel{(b)}{\ge} \E_{1}^{(s^\star)} \left[ \Lambda^{(s^\star)}(\Yv_j|\xv_j^{(s^\star)}) \right] - (|\Sc|-1)\rho^\eta \\
        &~~~ \times \left( \E_{1}^{(s^\star)} \left[ \Lambda^{(s^\star)}(\Yv_j|\xv_j^{(s^\star)}) \right] - \min_{s \ne s^\star} \E_{1}^{(s^\star)} \left[ \Lambda^{(s)}(\Yv_j|\xv_j^{(s)}) \right] \right) \\
		&\stackrel{(c)}{=} L  D( p_{Y|X}^{(s^\star)} \| p_{Y|X}^{(0)} | p_X^{(s^\star)} ) - (|\Sc|-1)\rho^\eta  L \gamma.
	\end{align*}
}
where $(a)$ follows since conditioned on $\Sh_j = s$, the likelihood ratio is independent of the past subblocks, $(b)$ is by Lem.~\ref{lem:MLE_error_conditional}, and $(c)$ follows from two facts that
\begin{align*}
		\E_{1}^{s^\star} \left[ \Lambda^{(s^\star)}(\Yv_j|\xv_j^{(s^\star)})\right] &= D( p^{(s^\star)} (\Yv_j|\xv_j^{(s^\star)}) \| p^{(0)} (\Yv_j|\xv_j^{(s^\star)}) ) \\
		&= L D( p^{(s^\star)} \| p^{(0)} | p_{X|s^\star} )
\end{align*}
and that the log-likelihood ratio of a single symbol is bounded above by $\gamma$. This proves the first claim of the lemma. The second claim is immediate by normalizing both sides by $L+1$ and considering large enough $L$ and $\eta$.
\end{IEEEproof}

We now address the final component required to combine all the intermediate bounds and conclude the achievability proof. To this end, we extend \cite[Thm. 2]{Veeravalli--Fellouris--Moustakides2024} to the JCCS strategy. 
\begin{theorem}\label{thm:Nb_bound}
	Suppose that the true state is $s$. Then, as $\eta, L, b \to \infty$ as in Tab.~\ref{tab:scaling},
	\begin{align*}
		\sup_{x^w} \E_{1}^{(s)} [ N_{b} | x^w] \le \frac{b}{D( p^{(s)} \| p^{(0)} | p_{X|s} )} ( 1+o(1) )
	\end{align*}
\end{theorem}
\begin{IEEEproof}
Instead of our CuSum test, consider a slightly different test:
\begin{align*}
	\Wv_j' &= \begin{cases} 
            0 & \text{if } j \le 2\eta \\
	 \Wv_{j-1}' + \Lambda^{(\Sh_j)}(\Yv_j|\xv_j) & \text{if } j > 2\eta,
    \end{cases}
\end{align*}
and use the following stopping rule $\Nv_{b}' := \inf \{j > 2\eta : \Wv_{j}' \ge b\}$. Note that its symbolwise translation is $N_{b}' = (L+1) \Nv_{b}'$. Since $\Wv_{j}'$ does not take $\max\{\cdot, 0\}$ compared to $\Wv_{j}$ in its definition, we have the relation $\Wv_{j} \ge \Wv_{j}'$ and $\Nv_{b} \le \Nv_{b}'$. Then, it is sufficient to show that for any fixed $x^w$,
\begin{align*}
	\E_1^{(s)} [N_{b}' | x^w] \le \frac{b}{D( p^{(s)} \| p^{(0)} | p_{X|s} )} ( 1+o(1) )
	\end{align*}
since the right side is independent of $x^w$.
	
To show this, we apply~\cite[Proposition 1]{Veeravalli--Fellouris--Moustakides2024} by substituting $n \leftarrow j$, $w \leftarrow \eta$, $P \leftarrow \P_1^{s}(\cdot | x^w)$, $\Gc_j \leftarrow \Fc_j$,$Y_j \leftarrow \Lambda^{(\sh_j)}(\Yv_j)$, $\tau_b \leftarrow \Nv_{b}'$. Also, we substitute 
\begin{align*}
	\mu_* &\leftarrow L \cdot D( p^{(s^\star)} \| p^{(0)} | p_{X^|s^\star} ) - L \gamma (|\mathcal{S}|-1) \rho^\eta,  ~~ \text{by Lem.~\ref{lem:drift_LB}}, \\
	\mu_0 & \leftarrow L \cdot D( p^{(s^\star)} \| p^{(0)} | p_{X|s^\star} ).
\end{align*}
Then, for $\eta L = o(b)$, \cite[Proposition 1]{Veeravalli--Fellouris--Moustakides2024} gives
\begin{align*}
	&\E[\tau_b] \le \frac{ b + \eta \mu + C(1 + \sqrt{b} + \sqrt{\eta}) }{\mu_*} \\
	\Rightarrow ~ &\E_1^{(s)}[\Nv_{b}' | x^w ] \\
    &\le \frac{b + \eta L D( p^{(s^\star)} \| p^{(0)} | p_{X|s^\star} ) + C(1 + \sqrt{b} + \eta) }{L D( p^{(s^\star)} \| p^{(0)} | p_{X|s^\star} ) - L \gamma (|\mathcal{S}|-1) \rho^\eta} \\
	&= \frac{b}{L D( p^{(s^\star)} \| p^{(0)} | p_{X|s^\star} )} (1+o(1)).
\end{align*}
As $\Nv'_{b+\gamma L}$ is the subblock index, converting it to a symbol index yields
\begin{align*}
	\E_{1}^{(s)} [ N_{b} | x^w] &\le (L+1) \E_1^{(s)}[\Nv_{b}' | x^w ] \\
	&\le \frac{(L+1)b}{L D( p^{(s^\star)} \| p^{(0)} | p_{X|s^\star} )} (1+o(1)) \\
	&= \frac{b}{D( p^{(s^\star)} \| p^{(0)} | p_{X|s^\star} )} (1+o(1))
\end{align*}
when $L \to \infty$.
\end{IEEEproof}

\section{Numerical Examples}\label{sec:examples}

\subsection{Binary channel example}~\label{sec:BIBO_ex}
\begin{figure}[t]
	\centering
        \resizebox{!}{8em}{\begin{tikzpicture}[node distance=.6cm and 1cm, start chain]

  \node (input0) {$0$};
  \node[below=1.5cm of input0] (input1) {$1$};
  \node[right=2.5cm of input0] (output0) {$0$};
  \node[right=2.5cm of input1] (output1) {$1$};

\draw[-] (input0.east) -- node[above] {$1-\epsilon_0$} (output0.west);
\draw[-] (input1.east) -- node[below] {$1-\epsilon_{1}$} (output1.west);
\draw[-] (input0.east) -- node[above, xshift=0.5cm] {$\epsilon_0$} (output1.west);
\draw[-] (input1.east) -- node[below, xshift=0.5cm] {$\epsilon_1$} (output0.west);
\end{tikzpicture}}
	\caption{Binary-input binary-output channel with crossover probabilities $\e_0$, $\e_1$.}
	\label{fig:binary_example}
\end{figure}
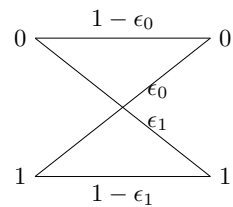
\begin{figure}[t]
\centering
\resizebox{.45\textwidth}{!}{\input{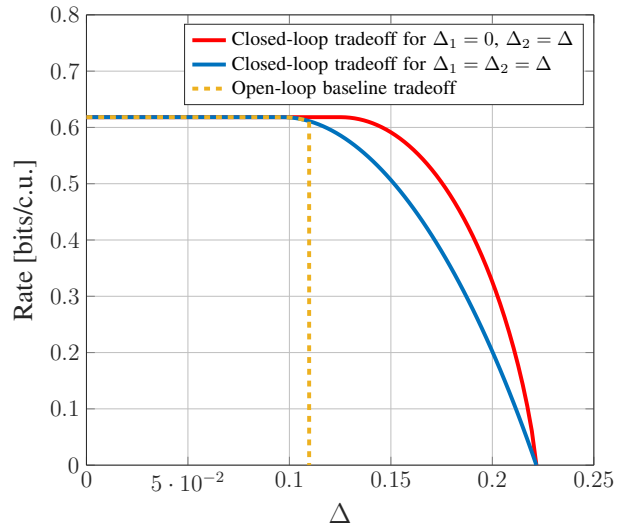}}
\caption{Achievable region for the BIBO channel example.}
\label{fig:achievable_region}
\end{figure}
Consider a binary-input binary-output (BIBO) channel with crossover probabilities $\e_0$, $\e_1$ as shown in Fig.~\ref{fig:binary_example}. Suppose that the communication channel is state-independent and has parameters $\epsilon_0 = 0$ and $\epsilon_1 = 0.2$, representing a $Z$-channel with crossover probability $0.2$:
\begin{align*}
	&p_{\Yt | X} (0|0) = 1, \quad p_{\Yt | X} (1|0) = 0, \\
	&p_{\Yt | X} (0|1) = 0.2, \quad p_{\Yt | X} (1|1) = 0.8.
\end{align*}
The sensing channel operates in three possible states. In the base state ($S=0$), the channel is modeled as a binary symmetric channel (BSC) with a crossover probability of $0.1$, i.e., $\epsilon_0 = \epsilon_1 = 0.1$. For state $S=1$, the sensing channel is a BIBO channel with crossover probabilities $\epsilon_0 = 0.1$ and $\epsilon_1 = 0.3$. Similarly, for state $S=2$, the sensing channel is a BIBO channel with crossover probabilities $\epsilon_0 = 0.3$ and $\epsilon_1 = 0.1$.

As a baseline comparison, we consider the achievable rate-delay pair with the open-loop strategy in Thm.~\ref{thm:open-loop}. To simplify expressions, let $\alpha := p_X(1)$. Then, the mutual information term is given by  
\begin{align*}
	I(X;\Yt) &= H(\Yt) - H(\Yt|X) \\
	&= H_2(0.8\alpha) - \alpha H_2(0.2),
\end{align*}
where $p_{\Yt}(1) = 0.8\alpha$ and $H_2(\cdot)$ is the binary entropy function.
The KL divergences are calculated as
\begin{align*}
	D( p^{(1)} \| p^{(0)} | p_X) &\approx 0.2217\alpha,\\
	D( p^{(2)} \| p^{(0)} | p_X) &\approx 0.2217(1-\alpha).
\end{align*}
With the expressions at hand, Fig.~\ref{fig:achievable_region} compares the baseline open-loop tradeoff in Thm.~\ref{thm:open-loop} and our closed-loop tradeoff given in Thm.~\ref{thm:theorem1}.   
As the channel is a $Z$-channel, the data rate exhibits asymmetry with respect to $\alpha = p_X(1)$. However, the KL divergences are reversely symmetric. In the baseline strategy, the mutual information is evaluated against the minimum of the two reversely symmetric KL divergences, resulting in the most inner curve (yellow) in the figure. 
The closed-loop strategy outlined in Thm.~\ref{thm:theorem1} yields superior tradeoff performances. Specifically, the second outer (blue) curve represents the tradeoff for the case $\Delta_1 = \Delta_2 = \Delta$, while the most outer (red) curve corresponds to the case $\Delta_1 = 0$ and $\Delta_2 = \Delta$. We note that the most outer and second outer curves also correspond to the specialized tradeoffs when the state space is binary, i.e., $\Sc = \{1\}$ and $\Sc = \{2\}$, respectively. Since Thm.~\ref{thm:theorem1} aligns with these specialized tradeoffs, it highlights the adaptive capability of the closed-loop strategy.

\subsection{MIMO Gaussian example}

\begin{figure}[t]
\begin{center}
\resizebox{.65\textwidth}{!}{\input{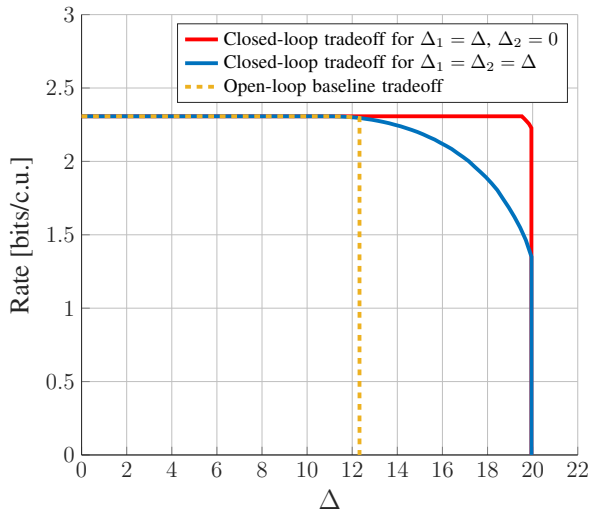}}
\caption{Achievable region for the Gaussian MIMO channel example.}
\label{fig:mimo_achievable_region}
\end{center}
\end{figure}

Next, we explore two MIMO Gaussian examples. The first one is a multiple post-state example, where the MIMO Gaussian communication channel is given by
\begin{align*}
    \Yt_i =\begin{cases}
    \Ht_0 \xv_i + \tilde{\Zv}_i &\quad\text{ for } i<\nu\\
    \Ht_1 \xv_i + \tilde{\Zv}_i &\quad\text{ for } i\ge\nu, S=1\\
    \Ht_2 \xv_i + \tilde{\Zv}_i &\quad\text{ for } i\ge\nu, S=2
    \end{cases}
\end{align*}
and the MIMO echo channel is given by
\begin{align*}
    Y_i =\begin{cases}
    \Zv_i &\quad\text{ for } i<\nu\\
    H_1\xv_i + \Zv_i &\quad\text{ for } i\ge\nu, S=1\\
    H_2\xv_i + \Zv_i &\quad\text{ for } i\ge\nu, S=2,
    \end{cases}
\end{align*}
where $\Zv_i$ and $\tilde{\Zv}_i$ are independent vector Gaussian noise with $\Nc(0,I)$ and the channel inputs are subject to a power constraint $\sum_{i=1}^n \|\xv_i\|^2 \le nP$. Compared to the scalar Gaussian case, which will be stated shortly, there are several interesting tradeoff relations involved depending on the channel gain matrices $H_i$'s and $\Ht_i$'s.

\begin{theorem}\label{thm:mimo}
For a post-state set $\Sc$, a rate-delay pair $(R, \Delta)$ is achievable for some $\{\Nc(0, \Sigma_{\Xv_s}): s\in\Sc\}$ if, 
\begin{align}
	R &< \min_s \frac{1}{2}\log\left|I+\tilde{H}_s\Sigma_{\Xv_s}\tilde{H}_s^T\right|, \label{eq:mimo_rate}\\
	\Delta_{s} &< \frac{1}{2}\tr\left(\Lambda_s \Sigma_{\bar X}\right), \label{eq:mimo_delta}
\end{align}
for all $s\in\Sc$, where $\Gamma_s = {H}_s^T{H}_s$, $\Gamma_s = U_s\Lambda_sU_s^T$ is the SVD of $\Gamma$, and $\bar\Xv_s = U_s^T\Xv_s$.
\end{theorem}
The above theorem is a direct application of Thm.~\ref{thm:theorem1} evaluated with vector Gaussian distributions $\{\Nc(0, \Sigma_{\Xv_s}): s\in\Sc\}$.

For an example evaluation, we assume that the power constraint is $P=10$ and the channels are given by
\begin{align}
    H_1 = \begin{bmatrix}
        2 & 0 \\
        0 & 1 
    \end{bmatrix},\quad  \tilde{H}_1 = \begin{bmatrix}
        1 & 0 \\
        1  & -1 
    \end{bmatrix},
\end{align}
and
\begin{align}
    H_2 = \begin{bmatrix}
        1 & 0 \\
        0 & 2 
    \end{bmatrix},\quad  \tilde{H}_2 = \begin{bmatrix}
        1/\sqrt{2} & 1/\sqrt{2} \\
        1 & 0 
    \end{bmatrix}.
\end{align}

In Fig.~\ref{fig:mimo_achievable_region}, we compare the performance of the proposed closed-loop tradeoff with the baseline open-loop strategy in Thm.~\ref{thm:open-loop}. 
The baseline strategy's tradeoff is given by the most inner curve (yellow) in the figure. 
The closed-loop strategy in Thm.~\ref{thm:theorem1} demonstrates superior tradeoff performances. Specifically, the second outer curve (blue) represents the tradeoff for the case $\Delta_1 = \Delta_2 = \Delta$, while the most outer (red) curve corresponds to the case $\Delta_1 = 0$ and $\Delta_2 = \Delta$.

\begin{remark} 
For the scalar Gaussian model, $X\sim \Nc(0,P)$ attains the rate-delay pair
\begin{align*}
    \left(\frac{1}{2}\log\left(1+\min\{\tilde{h}^2_1, \tilde{h}^2_2\}P\right), \frac{h^2_{s^\star}}{2}P\right) = (C, \bar\Delta)
\end{align*}
when the true change state is $s^\star\in\{1,2\}$. Note that this point is the optimal tradeoff in this case since $X\sim \Nc(0,P)$ simultaneously attains the capacity $C$ when $\Delta=\mathbf{0}$, and the optimal delay $\bar\Delta$ when $R=0$.
\end{remark}

\begin{figure}[t]
\begin{center}
\includegraphics[width=0.5\textwidth]{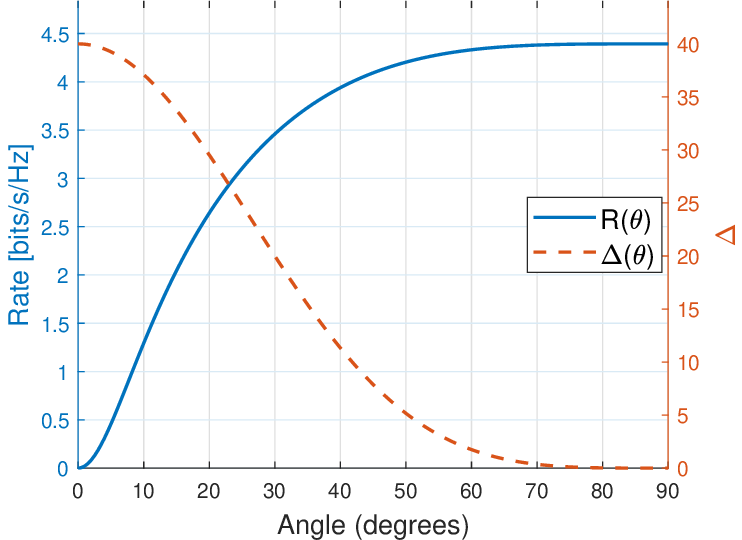}
\caption{Achievable rate and $\Delta$ vs. beamforming angle for the single post-state MIMO example.}
\label{fig:mimo_tradeoff}
\end{center}
\end{figure}

Next, we consider a single post-state MIMO example to more explicitly highlight the tradeoff between communication and QCD without the additional complexity arising from multiple states. We adopt a scenario with a multi-antenna transmitter with $M$ antennas and a single-antenna communication receiver. Following the formulation as in~\cite{Zheng--Lops--Eldar--Wang2019}, the MIMO Gaussian sensing channel, i.e., for QCD, is modeled similar to the previous example with
\begin{align*}
    Y_i = \begin{cases}
        \Zv_i, & \text{for } i < \nu, \\
        H_1 \xv_i + \Zv_i, & \text{for } i \ge \nu,
    \end{cases}
\end{align*}
where $H_1 = \av_r(\theta_1) \av_t(\theta_1)^H$, and $\av_t(\cdot), \av_r(\cdot)$ denote the transmit and receive steering vectors of a uniform linear array (ULA) with half-wavelength antenna spacing and angle of arrival (AoA) $\theta_1 \in [0, \pi/2]$, i.e., 
\begin{align*}
    \av(\theta) = 
    \begin{bmatrix}
        1 & e^{j\pi\sin(\theta)} & \ldots & e^{j\pi(M-1)\sin(\theta)}
    \end{bmatrix}^T
\end{align*}
The communication channel has no state and is modeled as
\begin{align*}
    \Yt_i = \Ht \xv_i + \tilde{Z}_i,
\end{align*}
where $\Ht =  \av_t(\theta_c)^H$, and $\av_t(\theta_c)$ is the transmit steering vector for the communication receiver located at AoA $\theta_c \in [0, \pi/2]$.
The transmit signal is set as $\xv_i = \xv_i(\theta) = \vv(\theta) x_i$, where $\vv(\theta)$ is a unit-norm beamforming vector in direction $\theta$, and $x^n = (x_1, \ldots, x_n)$ is an i.i.d. Gaussian codeword with average power $P$.

For our evaluation, we choose $\theta_c = \pi/2$, $\theta_1 = 0$, $P = 10$, and $M=2$. 
We evaluate the achievable rate $R(\theta)$ and the delay sensitivity bound $\Delta(\theta)$ for each beamforming direction $\theta$ as specified by Thm.~\ref{thm:mimo}. 
The results, shown in Fig.~\ref{fig:mimo_tradeoff}, demonstrate the fundamental tradeoff between communication rate and detection sensitivity. For example, beamforming toward the communication receiver (at $\theta_c$) maximizes the communication rate but yields the worst detection sensitivity, while beamforming toward the QCD target (at $\theta_1$) achieves the best detection performance but zero communication rate.
By steering the beam between these two directions, the transmitter can adjust the balance between communication and sensing performance, demonstrating the inherent tradeoff in this joint design problem.

\section{Discussion}
In this work, we analyzed the fundamental tradeoff between communication rate and quickest change detection (QCD) delay in integrated sensing and communication (ISAC) systems under a monostatic setup. We proposed a novel feedback-based joint communication and QCD subblock coding strategy (JCCS) and characterized the achievable rate-delay region using state-dependent mutual information and KL divergence. Additionally, we established a converse result, proving the asymptotic optimality of the proposed detection algorithm for the JCCS framework. Illustrative examples for various channel models highlighted the practical insights of our approach. By addressing key technical challenges, this work provides a foundational understanding and practical guidelines for designing ISAC systems with optimal communication and sensing performance.

\appendices 

\section{Proof of Theorem \ref{thm:conv}} \label{app:conv_proof}

For the converse proof, we need the following lemma and then apply Lai's theorem~\cite[Thm.~1]{Lai1998}.
\begin{lemma}\label{eq:ml_lai_cond}
For every $\delta>0$ and message $m\in[1:2^{nR}]$    
\begin{align*}
    \sup_{j_\nu} \esssupe \P_\nu^{(s)}\Bigg(&\max_{1\le t\le k} \sum_{i=j_\nu+1}^{j_\nu+t} \Lambda^{(s)}(\Yv_j|\Xv_j) \\
    &> L \cdot D^{(s)}_{p(x|s)}(1+\delta)k \Big|~\Fc_{\nu-1}(m)\Bigg)
\end{align*}
tends to zero as $k\to\infty$, where $D^{(s)}_{p_X} = D(p^{(s)}\| p^{(0)}|p_X)$.
\end{lemma}
If the above lemma holds, the proof of which will be discussed later, then by~\cite[Thm.~1]{Lai1998}, for any block-wise stopping rule $\Nv$, we have
\begin{align*}
&\sup_{j_\nu \ge 1} \esssupe \E_{\nu}^{(s)} (\Nv-j_\nu|\Fc_{\nu - 1}) \\
&\quad \ge |\log \alpha|\left( {\frac{1}{L\cdot D(p^{(s)}\| p^{(0)}|p(x|s))} + o(1)} \right) ~ \text{as $\alpha \to 0$.}
\end{align*}

Next, we translate the above block-wise delay bound to symbol-wise delay using the following bound: 
\begin{align*}
\Nv-j_\nu 
&= \frac{1}{(L+1)}\left(N-(L+1)j_\nu\right)\\
&\le \frac{1}{(L+1)}\left(N-\nu+ 1\right),
\end{align*}
since $(L+1)\Nv = N$ and $\nu \le j_\nu(L+1)$.
Overall, we have
\begin{align*}
&\sup_{\nu \ge 1} \esssupe \E_{\nu}(N-\nu+1|\Fc_{\nu - 1}) \\
&\quad \ge |\log \alpha|\left( {\frac{1}{D(p^{(s)}\| p^{(0)}|p(x|s))} + o(1)} \right) ~ \text{as $\alpha \to 0$.}
\end{align*}
which concludes the proof. The rest of this section is devoted to the proof of Lem.~\ref{eq:ml_lai_cond}.

\begin{IEEEproof}[Proof of Lem.~\ref{eq:ml_lai_cond}]
Assume $S_i=s$ for $i\ge \nu$. Let $\pi(q)=p(x|q)$ be the type of the subblock $\xv_j(m|q)=(x_{j1},\ldots, x_{jL})$ in the codebook $\Cc_{JCCS}^{(n)}$. Then, we have the relation
\begin{align}
    \frac{1}{L}\sum_{i=1}^L D(p^{(s)}(y|x_{ji})\| p^{(0)}(y|x_{ji})) = D^{(s)}_{\pi(q)} 
\end{align}
where $D^{(s)}_{p_X} = D(p^{(s)}\| p^{(0)}|p_X)$. Then, 
\begin{align}
    &\P_\nu^{(s)}\Bigg(\max_{1\le t\le k} \sum_{j=j_\nu+1}^{j_\nu+t} \Lambda^{(s)}_j > LD^{(s)}_{p(x|s)}(1+\delta)k \Big| ~\Fc_{\nu-1}(m)\Bigg) \nn \\
    &\le \P_\nu^{(s)}\Bigg(\max_{1\le t\le k} \sum_{j=j_\nu+1}^{j_\nu+t} \Big(\Lambda^{(s)}_j -LD^{(s)}_{\pi(\Sh_j)}\Big)>  \nn\\
    &\Big(kLD^{(s)}_{p(x|s)} - \sum_{j=1}^k LD^{(s)}_{\pi(\Sh_j)}\Big)+LD^{(s)}_{p(x|s)}\delta k  \Big| ~\Fc_{\nu-1}(m) \Bigg) \nn \\
    &\le \P_\nu^{(s)} \Bigg( \max_{1\le t\le k} \chi_t > 
    \frac{L}{k} \sum_{j=j_\nu + 1}^{j_\nu + \eta} \Delta^{(s)}_j \nn\\
    &~~~~~ + \frac{L}{k} \sum_{j=j_\nu + \eta + 1}^{j_\nu + k} \Delta^{(s)}_j+ LD^{(s)}_{p(x|s)}\delta \Big| ~\Fc_{\nu-1}(m) \Bigg) \label{eq:long_eq}
\end{align}
where $\Lambda^{(s)}_j := \Lambda^{(s)}(\Yv_j|\Xv_j(\Sh_j))$, $\Delta^{(s)}_j :=  D^{(s)}_{p(x|s)} - D^{(s)}_{\pi(\Sh_j)}$, and 
$\chi_t := \frac{1}{k}\sum_{j=j_\nu+1}^{j_\nu+t} \left( \Lambda^{(s)}_j - LD^{(s)}_{\pi(\Sh_j)}\right)$.
Note that the first term in the right side can be made arbitrarily small since $| \Delta^{(s)}_j | < c$ for some absolute constant $c>0$ and $L\eta = o(k)$. Hence, for any $\epsilon > 0$, one can take large enough $k, L$ in our asymptotic regime such that
$ \frac{L}{k} \sum_{j=j_\nu + 1}^{j_\nu + \eta} \Delta^{(s)}_j > - \epsilon$.
Take $\epsilon = L D^{(s)}_{p(x|s)} \delta/2$. Then, \eqref{eq:long_eq} can be bounded by $\P_\nu^{(s)} (\Ac | \Fc_{\nu-1}(m))$ where
\begin{align*}
\Ac := \Bigg\{& \max_{1\le t\le k} \chi_t> \frac{1}{k} \sum_{j=j_\nu + \eta + 1}^{j_\nu + k}  L\Delta_j^{(s)} + \frac{LD^{(s)}_{p(x|s)}\delta}{2} \Bigg\}.
\end{align*}
Next, define the event $\Bc_s = \{\Sh_j = s \text{ for all } j \in [j_\nu+\eta+1:j_\nu+k]\}$. Then,
\begin{align*}
\P_\nu^{(s)} (\Ac | \Fc_{\nu-1}(m)) &= \P_\nu^{(s)} ( \Bc_s ) \P_\nu^{(s)} (\Ac | \Fc_{\nu-1}(m), \Bc_s)\\
&\quad+ \P_\nu^{(s)} ( \Bc_s^c ) \P_\nu^{(s)} (\Ac | \Fc_{\nu-1}(m), \Bc_s^c) \\
&\le \P_\nu^{(s)} (\Ac | \Fc_{\nu-1}(m), \Bc_s) + \P_\nu^{(s)} ( \Bc_s^c ) \\
&\le \P_\nu^{(s)} (\Ac | \Fc_{\nu-1}(m), \Bc_s) + k (|\Sc|-1)\rho^{\eta},
\end{align*}
where the last inequality is due to Lem.~\ref{lem:MLE_error_unconditional}, and the second term tends to zero in our asymptotic regime with $\log k = o(\eta)$. To show that the first term vanishes as well, note that $\chi_t|\Bc_s, \Fc_{\nu-1}(m)$ is a martingale with respect to the filtration $\Gc_t = \sigma(\breve{Y}_1,\ldots, \breve{Y}_{j_\nu+t})$: 
\begin{align*}
    &\E^{(s)}_\nu\left[\chi_t | \Gc_{t-1}, \Fc_{\nu-1}, \Bc_s \right] = \E^{(s)}_\nu\left[\chi_t | \Gc_{t-1}, \Bc_s \right] \\
    &= \E^{(s)}_\nu \left[ \frac{1}{k}\sum_{j=j_\nu+1}^{j_\nu+t} \left( \Lambda^{(s)}_j - LD^{(s)}_{\pi(\Sh_j)} \right) \bigg| \Gc_{t-1}, \Bc_s \right] \\
    &= \chi_{t-1} + \frac{1}{k} \E^{(s)}_\nu\left[ \Lambda^{(s)}_{j_\nu+t} - LD^{(s)}_{\pi(\Sh_j)} \Big| \Gc_{t-1}, \Bc_s \right] = \chi_{t-1},
\end{align*}
where the first equality is due to the fact that $\chi_t\to (\Gc_{t-1}, \Bc_s)\to \Fc_{\nu-1}$ since $\chi_t$ is a function of code subblocks $(\xv_j, \yv_j)$ after time $\nu$, and the estimate $\Sh_j$ is a function of pilot observations that are conditioned by $\Gc_t$.
Then, since $(\cdot)^2$ is a convex function, $\chi_t^2$ is a submartingale~\cite[Thm.~4.2.6]{Durrett2019}. Thus, 
\begin{align*}
    &\P_\nu^{(s)} (\Ac | \Fc_{\nu-1}(m), \Bc_s) \\
    &\stackrel{(a)}{=} \P_\nu^{(s)} \Big(\max_{1\le t\le k} \chi_t > LD^{(s)}_{p(x|s)} \delta/2 \Big| \Fc_{\nu-1}(m), \Bc_s\Big) \\
    &\le \P_\nu^{(s)} \left(\max_{1\le t\le k} \chi_t^2 > \left( LD^{(s)}_{p(x|s)} \delta/2 \right)^2 \bigg| \Fc_{\nu-1}(m), \Bc_s\right) \\
    &\stackrel{(b)}{\le} \frac{\E_\nu^{(s)} \left[\chi_k^2|\Fc_{\nu-1}(m), \Bc_s\right]}{(LD^{(s)}_{p(x|s)} \delta/2)^2} \\
    &\le \frac{kLV}{(LD^{(s)}_{p(x|s)} \delta k/2)^2} = \Oc \left( \frac{1}{kL} \right) \to 0,
\end{align*}
where $(a)$ is due to $\pi(\Sh_j)=p(x|s)$ conditioned on $\Bc_s$, and $(b)$ is by Doob's submartingale inequality~\cite[Thm.~4.4.2]{Durrett2019}, and $V$ is a finite constant given in \eqref{eq:finite_V}.
\end{IEEEproof}

\section{Proof of Lemma~\ref{lem:MLE_error_unconditional}}\label{app: state_est_error_bound}

Extending \eqref{eq:def-LLR}, we define the $\ell$-th log-likelihood ratio between two states for pilot symbols
\begin{align*}
	\Lambda_\ell (s, s') := \log \frac{ p^{(s)}(\breve{y}_\ell | \breve{x}_\ell) }{ p^{(s')}(\breve{y}_\ell | \breve{x}_\ell) }.
\end{align*}
Then, by the definition of the MLE, $\sh_j = s^\star$ implies that for all $s' \ne s^\star$,
\begin{align*}
	\SUM_j(s^\star, s') := \sum_{\ell = j-\eta}^{j-1}\Lambda_\ell(s^\star, s') > 0.
\end{align*}
Therefore, for $j\ge j_\nu+\eta$,
\begin{align*}
\P_{\nu}^{(s^\star)} ( \Sh_j \ne s^\star ) &= \P_{\nu}^{(s^\star)} \left( \bigcup_{s' \ne s^\star} \{ \SUM_j(s^\star, s') \le 0 \} \right) \\
&\stackrel{(a)}{\le} \sum_{s' \ne s^\star} \P_{\nu}^{(s^\star)} \left( \SUM_j(s^\star, s') \le 0 \right) \\
&= \sum_{s' \ne s^\star} P_{\nu}^{(s^\star)} \left( \exp\left( -\frac{1}{2} \SUM_j(s^\star, s') \right) \ge 1 \right) \\
&\stackrel{(b)}{\le} \sum_{s' \ne s^\star} \E_{\nu}^{(s^\star)} \left[ \exp\left( -\frac{1}{2} \SUM_j(s^\star, s') \right) \right] \\
&\stackrel{(c)}{=} \sum_{s' \ne s^\star} \prod_{\ell=j-\eta}^{j-1} \E_{\nu}^{(s^\star)} \left[ \left( \frac{ p^{(s')}(\breve{Y}_\ell | \breve{X}_\ell) }{ p^{(s^\star)}(\breve{Y}_\ell | \breve{X}_\ell) } \right)^{1/2} \right] \\
&\stackrel{(d)}{=} \sum_{s' \ne s^\star} \rho(s', s^\star)^{\eta} \\
&\stackrel{(e)}{\le} \sum_{s' \ne s^\star} \rho^{\eta} = (|\Sc|-1) \rho^{\eta},
\end{align*}
where $(a)$ and $(b)$ are by the union bound and the Markov inequality, respectively, $(c)$ follows from the fact that the channel is memoryless, $(d)$ follows since pilot symbols are uniformly drawn, and $(e)$ is by the definition $\rho$, i.e., the maximum averaged Bhattacharya coefficient in~\eqref{eq:max_bhattacharya}.

\section{Proof of Vanishing entropy rate in \eqref{eq:ER1}} \label{app:entropy_rate}
Let $j_\nu$ be the subblock index that contains the change point $\nu$, i.e., $(j_\nu-1)(L+1) + 1 \le \nu \le j_\nu (L+1)$.
To show~\eqref{eq:ER1}, we can bound the entropy term by
\begin{align}
	&\frac{1}{k} H(\Sh^k) \le \frac{1}{k} \sum_{j=1}^k H(\Sh_j) \nn \\
    &= \frac{1}{k} \sum_{j=1}^{j_{\nu}+\eta-1} H(\Sh_j) + \frac{1}{k}\sum_{j=j_{\nu}+\eta}^{k} H(\Sh_j) \nn \\
	&\le \frac{|j_{\nu}+\eta-1|}{k} \log|\Sc| + \frac{1}{k}\sum_{j=j_{\nu}+\eta}^{k} H(\Sh_j). \label{eq:entropy_rate}
\end{align}
Let $I_j$ be the indicator random variable $I_j = \mathbf{1}\{\Sh_j=s^\star\}$ where $s^\star$ is the true change state. Then, for $j=j_{\nu}+\eta,\ldots, k$
\begin{align*}
	H(\Sh_j) &\le H(\Sh_j, I_j) = H(\Sh_j|I_j)+H(I_j)\\
	&= \P(I_j=1)H(\Sh_j|I_j=1)\\
    &\quad \quad \quad + \P(I_j=0)H(\Sh_j|I_j=0)+H(I_j)\\
	&\le \P(I_j=0)\log|\Sc|+H(I_j)\\
	&\stackrel{(a)}{\le} (|\Sc|-1) \rho^{\eta}\log|\Sc|+H(I_j),
\end{align*}
where step $(a)$ follows from Lem.~\ref{lem:MLE_error_unconditional}. Then, since $H(I_j)$ is the vanishing binary entropy function, \eqref{eq:entropy_rate} vanishes as well if $\eta = o(k)$. Note that this also implies that the fixed-length coding rate in~\cite[Thm.~1.3.1]{Han2003} tends to zero, since the spectral sup-entropy rate vanishes as a direct consequence of the convergence of the entropy rate and the Markov lemma.

\section{Proof of Theorem~\ref{thm:pure_comm}} \label{app:pure_comm_proof}
Assume $\nu=1$ and the post state is $s\in\Sc$. Then, by Fano's inequality
\begin{align*}
    nR &\le I(M;\Yt_s^n)+n\e_n\\
       &= \sum_{i=1}^n I(M;\Yt_{si}|\Yt_s^{i-1})+n\e_n\\
       &\le \sum_{i=1}^n I(M, Y^{i-1}_s, \Yt^{i-1}_s;\Yt_{si})+n\e_n\\
       &\stackrel{(a)}{=}\sum_{i=1}^n I(M, Y^{i-1}_s, X_{si}, \Yt^{i-1}_s;\Yt_{si})+n\e_n\\
       &\stackrel{(b)}{=} \sum_{i=1}^n I(X_{si};\Yt_{si})+n\e_n\\
       &\le nI(X_s; \Yt_{s})+n\e_n.
\end{align*}
where step $(a)$ follows since $X_{si}$ is a function of $(M, Y^{i-1}_s)$, and step $(b)$ follows from the Markov relation $(M, Y_s^{i-1}, \Yt_{s}^{i-1})\to X_{si} \to \Yt_{si}$.

Now moving forward to the  general case with $\nu\ge 1$,
\begin{align}
    nR &\le I(M;\Yt^n)+n\e_n\\
       &= I(M;\Yt_{\nu}^n)+I(M;\Yt^{\nu-1}|\Yt_{\nu}^n)+n\e_n.\label{eq:com_conv_ineq1}
\end{align}
Then,
\begin{align*}
    &I(M;\Yt^{\nu-1}|\Yt_\nu^n) \le I(M, \Yt_\nu^n;\Yt^{\nu-1})\\
    &= \sum_{i=1}^{\nu-1} I(M, \Yt_\nu^n; \Yt_i|\Yt^{i-1}) \le (\nu-1)I(X_{0}; \Yt_{0}),
\end{align*}
where we follow similar steps as in the $\nu=1$ case using the Markov relation $(M, Y_{0}^{i-1}, \Yt_{0}^{i-1}, \Yt^{n}_{\nu})\to X_{0,i}\to \Yt_{0,i}$ for $i=1,\ldots, \nu-1$. Thus, the term is $o(n)$ if $\max_{p_{X_{0}}}I(X_{0}; \Yt_{0}) < \infty$. On the other hand, the first term in~\eqref{eq:com_conv_ineq1} follows the same steps as the $\nu=1$ case, which gives the relation
\begin{align*}
    I(M;\Yt_{\nu}^n) \le (n-\nu+1)I(X_s; Y_s).
\end{align*}
Finally, by taking $n\to\infty$ we have 
\begin{align*}
    R\le \min_{s\in\Sc}\max_{p_{X_s}}I(X_s; Y_s).
\end{align*}

\end{document}